\documentclass[final,11pt,times,fleqn,hidelinks]{elsarticle}

\usepackage[utf8]{inputenc}
\usepackage[T1]{fontenc}
\usepackage[english]{babel}
\usepackage{amsmath}
\usepackage{amsfonts}
\usepackage{amssymb}
\usepackage{mathrsfs}
\usepackage{thmbox_aef}
\usepackage{array,multirow,tabularx}
\usepackage{xspace}
\usepackage{textcomp}
\usepackage{tikz}
\usetikzlibrary{decorations.pathreplacing,patterns}
\usepackage{titlesec}
\usepackage{caption} 
\usepackage{pdflscape}
\usepackage{hyperref}

\definecolor{turquoise}{RGB}{53, 123, 153} 
\definecolor{vert}{RGB}{194, 247, 50}

\definecolor{orange}{RGB}{250, 130, 1}
\colorlet{oranger}{orange!55!yellow}
\definecolor{jaune}{RGB}{255, 195, 0 }

\definecolor{brique}{RGB}{141, 48, 24}
\definecolor{brown}{RGB}{165, 137, 107}
\definecolor{chamois}{RGB}{200, 141, 75}

\definecolor{lilas}{RGB}{154, 107, 165}
\colorlet{parme}{magenta} 
\definecolor{mauve}{RGB}{161, 132, 220}

\definecolor{prunelle}{RGB}{106,24, 141}
\definecolor{prune}{RGB}{121, 7, 123}
\definecolor{violet}{RGB}{153, 0, 139}
\definecolor{violette}{RGB}{128, 0, 128} 

\definecolor{rose}{RGB}{255, 0, 127} 
\definecolor{framboise}{RGB}{141, 24, 59}

\colorlet{expli}{gray}

\newcommand{\dbloc}{$d$-block\xspace}
\newcommand{\dblocs}{$d$-blocks\xspace}
\newcommand{\Vshaped}{V-shaped\xspace}
\newcommand{\Vdbloc}{\Vshaped \dbloc}
\newcommand{\Vdblocs}{\Vshaped \dblocs}
\newcommand{\RSshaped}{$\rho$-$\sigma$-shaped\xspace}

\newcommand {\ie}{\textit{i.e.}\xspace}
\newcommand {\cf}{\textit{Cf.}\xspace}
\newcommand {\resp}{resp.\! }

\newcommand {\NN}{\mathbb{N}}
\newcommand {\NNe}{\NN^*}
\newcommand {\R}{\mathbb{R}}
\newcommand {\Rp}{\R_+}
\newcommand {\Rpe}{\R^*_+}

\newcommand{\J}{{J^<}}

\renewcommand {\=}{\!=\!}
\renewcommand {\-}{\!-\!}

\newcommand {\+}{\!+\!}
\newcommand {\iin}{\!\in\!}
\newcommand {\inclus}{\!\subseteq\!}

\renewcommand{\S}{\mathcal{S}}

\newcommand{\Part}{\vec{\mathcal{P}}_2^*(J)}
\newcommand{\PartC}{\vec{\mathcal{P}}_2(J)}

\newcommand{\lepb}{UCDDP\xspace}

\newcommand{\FET}{F^1}
\newcommand{\FDXf}{F^2}
\newcommand{\FIf}{F^{\,i}}
\newcommand{\FSf}{F^{\,s}}
\newcommand{\FISf}{F^{\,i + s}}
\newcommand{\FLP}{F\text{-}\textsc{lp}}
\newcommand{\FDXLP}{\FDXf\text{-}\textsc{lp}}
\newcommand{\FILP}{\FIf\text{-}\textsc{lp}}
\newcommand{\FSLP}{\FSf\text{-}\textsc{lp}}
\newcommand{\FISLP}{\FISf\text{-}\textsc{lp}}
\newcommand{\FDX}{F^2_l}
\newcommand{\FI}{F^{\,i}_l}
\newcommand{\FS}{F^{\,s}_l}
\newcommand{\FIS}{F^{\,i + s}_l}
\newcommand{\FF}{F^2_{d}}
\newcommand{\FFI}{F_{d}^{\,i}}
\newcommand{\FFS}{F_{d}^{\,s}}
\newcommand{\FFIS}{F_{d}^{\,i + s}}
\newcommand{\FRN}{F\text{-}\textsc{rn}}
\newcommand{\FDXRN}{\FDX\text{-}\textsc{rn}}
\newcommand{\FIRN}{\FI\text{-}\textsc{rn}}
\newcommand{\FSRN}{\FS\text{-}\textsc{rn}}
\newcommand{\FISRN}{\FIS\text{-}\textsc{rn}}
\newcommand{\FFRN}{\FF\text{-}\textsc{rn}}
\newcommand{\FFIRN}{\FFI\text{-}\textsc{rn}}
\newcommand{\FFSRN}{\FFS\text{-}\textsc{rn}}
\newcommand{\FFISRN}{\FFIS\text{-}\textsc{rn}}
\newcommand{\FA}{$\HRp$\xspace}
\newcommand{\FB}{$\HRRp$\xspace}
\newcommand{\FC}{$\FFIS\text{-}5\%$\xspace}
\newcommand{\FD}{$\FFIS$\xspace}

\newcommand{\HBF}{BF}
\newcommand{\HBFp}{BF+}
\newcommand{\HR}{R1}
\newcommand{\HRp}{R1+}
\newcommand{\HRR}{R2}
\newcommand{\HRRp}{R2+}
\newcommand{\LSproc}
{\texttt{Insert\_swap\_improvement}\xspace}

\newcommand{\nd}{{\footnotesize\#}nd}
\newcommand{\opt}{{\footnotesize\#}opt}
\renewcommand{\time}{time\xspace}
\newcommand{\LB}{L-gap\xspace}
\newcommand{\UB}{U-gap\xspace}

\newcommand{\BB}{Branch-and-Bound\xspace}

\newcommand {\Du}{\Delta_u(E,T)}
\newcommand {\Dv}{\Delta_v(E,T)}
\newcommand {\Duv}{\Delta_{u,v}(E,T)}
\newcommand {\Dud}{\Delta_u(\delta)}

\newcommand {\Duvd}{\Delta_{u,v}(\delta)}

\newcommand{\Mu}{M_u}
\newcommand{\Mpu}{M'_u}
\newcommand{\Muv}{M_{u,v}}
\newcommand{\Mtuv}{\widetilde{M}_{u,v}}

\renewcommand {\a}{\alpha}
\renewcommand {\b}{\beta}
\newcommand {\Ab}{\bar{A}}
\newcommand {\Bb}{\bar{B}}
\newcommand {\Au}{A(u) \!\cap\!E}

\newcommand {\Abu}{\Ab(u) \!\cap\!E}

\newcommand {\Bv}{B(u) \!\cap\!T}

\newcommand {\Bbv}{\Bb(v) \!\cap\!T}

\newcommand {\Avs}{A(v)\!\setminus\!\{u\} \!\cap\!E}
\newcommand {\Av}{A(v)\!\cap\!E}
\newcommand {\Abv}{\Ab(v) \!\cap\!E}
\newcommand {\Bu}{B(u)
\!\cap\!T}
\newcommand {\Bbu}{\Bb(u) \!\cap\!T}
\newcommand {\AAbuv}{A(u) \!\cap\!\Ab(v) \!\cap\!E}
\newcommand {\AAbvu}{A(v) \!\cap\!\Ab(u) \!\cap\!E}
\newcommand {\BBbuv}{B(u) \!\cap\!\Bb(v) \!\cap\!T}
\newcommand {\BBbvu}{B(v) \!\cap\!\Bb(u) \!\cap\!T}
\biboptions{authoryear}

\begin{document}
\journal{EJOR}

\newcommand{\modifRun}[1]{#1}
\newcommand{\modifRdeux}[1]{#1}
\newcommand{\modifRtrois}[1]{#1}
\newcommand{\modifAEF}[1]{#1}

\newcommand{\Q}[1]{\textcolor{red}{[#1]}}
\newcommand{\AF}[1]{}
\newcommand{\corrige}[2]{
\textcolor{blue}{[\textcolor{expli}{#1}$\rightarrow$#2]}
}

\newcommand{\z}{0}

\newcommand*{\tacheCouleurP}[4]{
	\fill [rounded corners,#4](#3-#2,\z + 0.1) rectangle  (#3,\z + 0.6);
	\draw[rounded corners] (#3-#2,\z+0.1) rectangle  (#3,\z+0.6);
	\draw (#3-#2/2,\z+0.82) node {\textcolor{#4}{#1}};
}
\newcommand*{\tacheCouleurPbas}[4]{
	\fill [rounded corners,#4](#3-#2,\z + 0.1) rectangle  (#3,\z + 0.6);
	\draw[rounded corners] (#3-#2,\z+0.1) rectangle  (#3,\z+0.6);
	\draw (#3-#2/2,\z-0.2) node {\textcolor{#4}{#1}};
}

\newcommand*{\tacheCouleur}[4]{
	\draw[rounded corners,#4,line width=1.5pt] (#3-#2,\z+0.1) rectangle  (#3,\z+0.6);
	\draw (#3-#2/2,\z+0.82) node {\textcolor{#4}{#1}};
}

\newcommand{\OrdoCombAvUn}[1]{
\begin{tikzpicture}[scale=0.55,every node/.style={transform shape}]
\newcommand{\decalage}{-4}
\tikzstyle{MonStyle}=[rounded corners,pattern color=brown,pattern=crosshatch dots]

\fill[MonStyle] (-7,+0.1) rectangle  (-6,0.6);
\fill[MonStyle] (-6,+0.1) rectangle  (-5,0.6);
\fill[MonStyle] (-3.5,+0.1+\decalage) rectangle  (-4.5,0.6+\decalage);
\fill[MonStyle] (-4.5,+0.1+\decalage) rectangle  (-5.5,0.6+\decalage);

\draw[->,brown, line width =1pt] (-6.3,-0.3) -- (-4.8,0.7+\decalage);
\draw[->,brown, line width =1pt] (-5.7,-0.3) -- (-4.2,0.7+\decalage) node[fill=white ,midway, draw, rounded corners,dotted] {$A(v) \!\cap\! \Ab(u) \!\cap\!E$};

\begin{scope}//avant le swap
\newcommand{\retroT}{-8};
\draw(\retroT-2.6,0) node{$\S$};
\draw [-](\retroT-2.2,0)--(0,0);
\draw[expli, dashed] (0,0)--(0.6,0);

\foreach \x in {1,2,3}  \tacheCouleur{}{1}{\retroT+\x}{orange};
\tacheCouleur{}{1.5}{-8}{orange};
\if#10
\else
\draw [color=orange,decorate,decoration={brace,amplitude=6pt},line width=1.5pt](-9.4,0.75) -- (-5.1,0.75) node[midway,above=7pt]{$\Ab(u) \!\cap\!E$};
\fi

\tacheCouleur{}{1}{-2.5}{oranger};
\tacheCouleur{}{0.7}{-1.8}{oranger};
\tacheCouleur{}{0.8}{-1}{oranger};
\tacheCouleur{}{1}{0}{oranger};
\if#10
draw[color=oranger](-1.9,1) node{$A(u) \!\cap\!E$};
\else
\draw [color=oranger,decorate,decoration={brace,amplitude=6pt},line width=1.5pt](-3.4,0.75) -- (-0.1,0.75) node[midway,above=7pt]{$A(u) \!\cap\!E$};
\fi

\draw [color=chamois,decorate,decoration={brace,amplitude=6pt},line width=1.5pt](-0.1,-0.05) -- (-7,-0.05);
\draw[color=chamois](-3.8,-0.7) node{$A(v) \!\cap\!E$};

\draw [color=brique,decorate,decoration={brace,amplitude=6pt},line width=1.5pt](-7,-0.05) -- (-9.5,-0.05);
\draw[color=brique](-8.4,-0.7) node{$\Ab(v) \!\cap\!E$};


\draw[line width =1.5pt](0,1)--(0,-0.2);
\draw (0,0) node[below=4pt]{$d$} node{$|$};

\tacheCouleurP{$u$}{1.5}{-3.5}{olive}
\end{scope}

\begin{scope}[yshift=\decalage cm]//après le swap
\newcommand{\retroT}{-8};
\draw(\retroT-2.6,0) node{$\S'$};
\draw [-](\retroT-2.2,0)--(0,0);
\draw[expli, dashed] (0,0)--(0.6,0);

\foreach \x in {1,2}  \tacheCouleur{}{1}{-5.5+\x}{orange};
\tacheCouleur{}{1}{-7.5}{orange};
\tacheCouleur{}{1.5}{-8.5}{orange};
\if#10
\else
\draw [color=orange,decorate,decoration={brace,amplitude=6pt},line width=1.5pt](-9.9,0.75) -- (-3.6,0.75) node[midway,above=7pt]{$\Ab(u) \!\cap\!E \!\cup\!\{v\}$};
\fi

\tacheCouleur{}{1}{-2.5}{oranger};
\tacheCouleur{}{0.7}{-1.8}{oranger};
\tacheCouleur{}{0.8}{-1}{oranger};
\tacheCouleur{}{1}{0}{oranger};
\if#10
draw[color=oranger](-1.9,1) node{$A(u) \!\cap\!E$};
\else
\draw [color=oranger,decorate,decoration={brace,amplitude=6pt},line width=1.5pt](-3.4,0.75) -- (-0.1,0.75) node[midway,above=7pt]{$A(u) \!\cap\!E$};
\fi

\draw [color=chamois,decorate,decoration={brace,amplitude=6pt},line width=1.5pt](-0.1,-0.05) -- (-5.5,-0.05);
\draw[color=chamois](-2.8,-0.7) node{$A(v) \!\setminus\! \{u\} \!\cap\!E$};

\draw [color=brique,decorate,decoration={brace,amplitude=6pt},line width=1.5pt](-7.5,-0.05) -- (-9.9,-0.05);
\draw[color=brique](-8.75,-0.7) node{$\Ab(v) \!\cap\!E$};


\draw[line width =1.5pt](0,1)--(0,-0.2);
\draw (0,0) node[below=4pt]{$d$} node{$|$};

\tacheCouleurPbas{$v$}{2}{-5.5}{turquoise}
\end{scope}

\end{tikzpicture}
}

\newcommand{\OrdoCombAvDeux}[1]{
\begin{tikzpicture}[scale=0.55,every node/.style={transform shape}]
\newcommand{\decalage}{-4}
\tikzstyle{MonStyle}=[rounded corners,pattern color=brown,pattern=crosshatch dots]

\fill[MonStyle] (-2.5,+0.1) rectangle  (-1.8,0.6);
\fill[MonStyle] (-3.5,+0.1) rectangle  (-2.5,0.6);
\fill[MonStyle] (-3.8,+0.1+\decalage) rectangle  (-4.5,0.6+\decalage);
\fill[MonStyle] (-4.5,+0.1+\decalage) rectangle  (-5.5,0.6+\decalage);

\draw[->,brown, line width =1pt] (-3,-0.3) -- (-5,0.7+\decalage);
\draw[->,brown, line width =1pt] (-2.2,-0.3) -- (-4.2,0.7+\decalage) node[fill=white ,midway, draw, rounded corners,dotted] {$A(u) \!\cap\! \Ab(v) \!\cap\!E$};

\begin{scope}//avant le swap
\newcommand{\retroT}{-8};
\draw(\retroT-2.6,0) node{$\S$};
\draw [-](\retroT-2.2,0)--(0,0);
\draw[expli, dashed] (0,0)--(0.6,0);

\foreach \x in {1,2,3}  \tacheCouleur{}{1}{\retroT+\x}{orange};
\tacheCouleur{}{1.5}{-8}{orange};
\if#10
\else
\draw [color=orange,decorate,decoration={brace,amplitude=6pt},line width=1.5pt](-9.4,0.75) -- (-5.1,0.75) node[midway,above=7pt]{$\Ab(u) \!\cap\!E$};
\fi

\tacheCouleur{}{1}{-2.5}{oranger};
\tacheCouleur{}{0.7}{-1.8}{oranger};
\tacheCouleur{}{0.8}{-1}{oranger};
\tacheCouleur{}{1}{0}{oranger};
\if#10
draw[color=oranger](-1.9,1) node{$A(u) \!\cap\!E$};
\else
\draw [color=oranger,decorate,decoration={brace,amplitude=6pt},line width=1.5pt](-3.4,0.75) -- (-0.1,0.75) node[midway,above=7pt]{$A(u) \!\cap\!E$};
\fi

\draw [color=chamois,decorate,decoration={brace,amplitude=6pt},line width=1.5pt](-0.1,-0.05) -- (-1.8,-0.05);
\draw[color=chamois](-0.9,-0.7) node{$A(v) \!\cap\!E$};

\draw [color=brique,decorate,decoration={brace,amplitude=6pt},line width=1.5pt](-1.8,-0.05) -- (-9.5,-0.05);
\draw[color=brique](-6,-0.7) node{$\Ab(v) \!\cap\!E$};


\draw[line width =1.5pt](0,1)--(0,-0.2);
\draw (0,0) node[below=4pt]{$d$} node{$|$};

\tacheCouleurP{$u$}{1.5}{-3.5}{olive}
\end{scope}

\begin{scope}[yshift=\decalage cm]//après le swap
\newcommand{\retroT}{-8};
\draw(\retroT-2.6,0) node{$\S'$};
\draw [-](\retroT-2.2,0)--(0,0);
\draw[expli, dashed] (0,0)--(0.6,0);

\foreach \x in {1,2,3}  \tacheCouleur{}{1}{\retroT+\x-0.5}{orange};
\tacheCouleur{}{1.5}{-8.5}{orange};
\if#10
\draw[color=orange](-7.5,1) node{$\Ab(u) \!\cap\!E$};
\else
\draw [color=orange,decorate,decoration={brace,amplitude=6pt},line width=1.5pt](-9.9,0.75) -- (-5.6,0.75) node[midway,above=7pt]{$\Ab(u) \!\cap\!E$};
\fi%

\tacheCouleur{}{1}{-4.5}{oranger};
\tacheCouleur{}{0.7}{-3.8}{oranger};
\tacheCouleur{}{0.8}{-1}{oranger};
\tacheCouleur{}{1}{0}{oranger};
\if#10
draw[color=oranger](-1.9,1) node{$A(u) \!\cap\!E$};
\else
\draw [color=oranger,decorate,decoration={brace,amplitude=6pt},line width=1.5pt](-5.4,0.75) -- (-0.1,0.75) node[midway,above=7pt]{\hspace*{0.8cm}$A(u) \!\cap\!E \!\cup\!\{v\}$};
\fi%

\draw [color=chamois,decorate,decoration={brace,amplitude=6pt},line width=1.5pt](-0.1,-0.05) -- (-1.8,-0.05);
\draw[color=chamois](-0.9,-0.7) node{$A(v) \!\cap\!E$};

\draw [color=brique,decorate,decoration={brace,amplitude=6pt},line width=1.5pt](-3.8,-0.05) -- (-10,-0.05)node[midway, below=7pt]{$\Ab(v) \!\setminus\! \{u\} \!\cap\!E$};


\draw[line width =1.5pt](0,1)--(0,-0.2);
\draw (0,0) node[below=4pt]{$d$} node{$|$};

\tacheCouleurPbas{$v$}{2}{-1.8}{turquoise}
\end{scope}
\end{tikzpicture}
}

\newcommand{\OrdoCombRetardUn}[1]{
\begin{tikzpicture}[scale=0.55,every node/.style={transform shape}]
\newcommand{\decalage}{-4}
\tikzstyle{MonStyle}=[rounded corners,pattern color=lilas,pattern=crosshatch dots]
\fill[MonStyle] (0.7,+0.1) rectangle  (2,0.6);
\fill[MonStyle] (2,+0.1) rectangle  (3,0.6);
\fill[MonStyle] (2.2,+0.1+\decalage) rectangle  (3.5,0.6+\decalage);
\fill[MonStyle] (3.5,+0.1+\decalage) rectangle  (4.5,0.6+\decalage);

\draw[->,lilas, line width =1pt] (1.7,-0.3) -- (3.2,+0.7+\decalage);
\draw[->,lilas, line width =1pt] (2.3,-0.3) -- (3.8,+0.7+\decalage) node[fill=white ,midway, draw, rounded corners,dotted] {$B(v) \!\cap\! \Bb(u) \!\cap\!T$};


\begin{scope}//avant le swap
\draw(-1.6,0) node{$\S$};
\draw [->](-1,0)--(9.4,0);
\draw[expli, dashed] (0,0)--(-1.4,0);

\tacheCouleur{}{1}{0}{magenta};
\tacheCouleur{}{0.7}{0.7}{magenta};
\tacheCouleur{}{1.3}{2}{magenta};
\tacheCouleur{}{1}{3}{magenta};
\if#10
\draw[color=magenta](1,1) node{$B(v) \!\cap\!T$};
\else
\draw [color=magenta,decorate,decoration={brace,amplitude=6pt},line width=1.5pt](-0.9,0.75) -- (2.9,0.75) node[midway, above=7pt]{$B(v) \!\cap\!T$};
\fi%

\tacheCouleur{}{1}{6}{violet};
\tacheCouleur{}{1.5}{7.5}{violet};
\tacheCouleur{}{0.8}{8.3}{violet};
\tacheCouleur{}{0.8}{9.1}{violet};
\if#10
\draw[color=violet](7,1) node{$\Bb(v) \!\cap\! T$};
\else
\draw [color=violet,decorate,decoration={brace,amplitude=6pt},line width=1.5pt](5.1,0.75) -- (8.9,0.75) node[midway, above=7pt]{$\Bb(v) \!\cap\! T$};
\fi%

\draw [color=mauve,decorate,decoration={brace,amplitude=6pt}, line width=1.5pt](0.7,-0.05) -- (-0.9,-0.05) node[midway, below =7pt]{$B(u) \!\cap\! T$};

\draw [color=prunelle,decorate,decoration={brace,amplitude=6pt}, line width=1.5pt](9,-0.05) -- (0.7,-0.05)
node[midway, below =7pt]{$\Bb(u) \!\cap\! T$};
\if#10
\draw[color=violet](7,1) node{$\Bb(v) \!\cap\! T$};
\else
\draw [color=violet,decorate,decoration={brace,amplitude=6pt},line width=1.5pt](5.1,0.75) -- (8.9,0.75) node[midway, above=7pt]{$\Bb(v) \!\cap\! T$};
\fi

\draw[line width =1.5pt](-1,1)--(-1,-0.2);
\draw (-1,0) node[below=4pt]{$d$} node{$|$};

\tacheCouleurP{$v$}{2}{5}{turquoise}
\end{scope}

\begin{scope}[yshift=\decalage cm]//après le swap
\draw(-1.6,0) node{$\S'$};
\draw [->](-1,0)--(9.4,0);
\draw[expli, dashed] (0,0)--(-1.4,0);

\tacheCouleur{}{1}{0}{magenta};
\tacheCouleur{}{0.7}{0.7}{magenta};
\tacheCouleur{}{1.3}{3.5}{magenta};
\tacheCouleur{}{1}{4.5}{magenta};
\if#10
\else
\draw [color=magenta,decorate,decoration={brace,amplitude=6pt},line width=1.5pt](-0.9,0.75) -- (4.4,0.75) node[midway, above=7pt]{$B(v) \!\cap\!T\!\cup\!\{u\}$ \hspace*{0.4cm}};
\fi%

\tacheCouleur{}{1}{5.5}{violet};
\tacheCouleur{}{1.5}{7}{violet};
\tacheCouleur{}{0.8}{7.8}{violet};
\tacheCouleur{}{0.8}{8.6}{violet};
\if#10
\else
\draw [color=violet,decorate,decoration={brace,amplitude=6pt},line width=1.5pt](4.6,0.75) -- (8.4,0.75) node[midway, above=7pt]{$\Bb(v) \!\cap\! T$};
\fi%

\draw [color=mauve,decorate,decoration={brace,amplitude=6pt}, line width=1.5pt](0.7,-0.05) -- (-0.9,-0.05);
\draw[color=mauve](-0.1,-0.7) node{$B(u) \!\cap\! T$};

\draw [color=prunelle,decorate,decoration={brace,amplitude=6pt}, line width=1.5pt](8.5,-0.05) -- (2.1,-0.05)node[midway, below=7pt]{$\Bb(u) \!\cap\! T \!\setminus\!\{v\}$};


\draw[line width =1.5pt](-1,1)--(-1,-0.2);
\draw (-1,0) node[below=4pt]{$d$} node{$|$};

\tacheCouleurPbas{$u$}{1.5}{2.2}{olive}
\end{scope}
\end{tikzpicture}
}

\newcommand{\OrdoCombRetardDeux}[1]{
\begin{tikzpicture}[scale=0.55,every node/.style={transform shape}]
\newcommand{\decalage}{-4}
\tikzstyle{MonStyle}=[rounded corners,pattern color=lilas,pattern=crosshatch dots]
\fill[MonStyle] (5,+0.1) rectangle  (6,0.6);
\fill[MonStyle] (6,+0.1) rectangle  (7.5,0.6);
\fill[MonStyle] (3,+0.1+\decalage) rectangle  (4,0.6+\decalage);
\fill[MonStyle] (4,+0.1+\decalage) rectangle  (5.5,0.6+\decalage);

\draw[->,lilas, line width =1pt] (5.7,-0.3) -- (3.7,0.7+\decalage);
\draw[->,lilas, line width =1pt] (6.4,-0.3) -- (4.4,0.7+\decalage) node[fill=white ,midway, draw, rounded corners,dotted] {$B(u) \!\cap\! \Bb(v) \!\cap\!T$};


\begin{scope}//avant le swap
\draw(-1.6,0) node{$\S$};
\draw [->](-1,0)--(9.4,0);
\draw[expli, dashed] (0,0)--(-1.4,0);

\tacheCouleur{}{1}{0}{magenta};
\tacheCouleur{}{0.7}{0.7}{magenta};
\tacheCouleur{}{1.3}{2}{magenta};
\tacheCouleur{}{1}{3}{magenta};
\if#10
\draw[color=magenta](1,1) node{$B(v) \!\cap\!T$};
\else
\draw [color=magenta,decorate,decoration={brace,amplitude=6pt},line width=1.5pt](-0.9,0.75) -- (2.9,0.75) node[midway, above=7pt]{$B(v) \!\cap\!T$};
\fi%

\tacheCouleur{}{1}{6}{violet};
\tacheCouleur{}{1.5}{7.5}{violet};
\tacheCouleur{}{0.8}{8.3}{violet};
\tacheCouleur{}{0.8}{9.1}{violet};
\if#10
\draw[color=violet](7,1) node{$\Bb(v) \!\cap\! T$};
\else
\draw [color=violet,decorate,decoration={brace,amplitude=6pt},line width=1.5pt](5.1,0.75) -- (8.9,0.75) node[midway, above=7pt]{$\Bb(v) \!\cap\! T$};
\fi%

\draw [color=mauve,decorate,decoration={brace,amplitude=6pt}, line width=1.5pt](7.5,-0.05) -- (-0.9,-0.05);
\draw[color=mauve](3.5,-0.7) node{$B(u) \!\cap\! T$};

\draw [color=prunelle,decorate,decoration={brace,amplitude=6pt}, line width=1.5pt](9,-0.05) -- (7.5,-0.05);
\draw[color=prunelle](8.3,-0.7) node{$\Bb(u) \!\cap\! T$};


\draw[line width =1.5pt](-1,1)--(-1,-0.2);
\draw (-1,0) node[below=4pt]{$d$} node{$|$};

\tacheCouleurP{$v$}{2}{5}{turquoise}
\end{scope}

\begin{scope}[yshift=\decalage cm]//après le swap
\draw(-1.6,0) node{$\S'$};
\draw [->](-1,0)--(9.4,0);
\draw[expli, dashed] (0,0)--(-1.4,0);

\tacheCouleur{}{1}{0}{magenta};
\tacheCouleur{}{0.7}{0.7}{magenta};
\tacheCouleur{}{1.3}{2}{magenta};
\tacheCouleur{}{1}{3}{magenta};
\if#10
\else
\draw [color=magenta,decorate,decoration={brace,amplitude=6pt},line width=1.5pt](-0.9,0.75) -- (2.9,0.75) node[midway, above=7pt]{$B(v) \!\cap\!T $};
\fi%

\tacheCouleur{}{1}{4}{violet};
\tacheCouleur{}{1.5}{5.5}{violet};
\tacheCouleur{}{0.8}{7.8}{violet};
\tacheCouleur{}{0.8}{8.6}{violet};
\if#10
\else
\draw [color=violet,decorate,decoration={brace,amplitude=6pt},line width=1.5pt](3.1,0.75) -- (8.4,0.75) node[midway, above=7pt]{\hspace*{0.9cm}$\Bb(v) \!\cap\! T \!\cup\!\{u\}$};
\fi%

\draw [color=mauve,decorate,decoration={brace,amplitude=6pt}, line width=1.5pt](5.5,-0.05) -- (-0.9,-0.05);
\draw[color=mauve](2.5,-0.7) node{$B(u) \!\cap\! T \!\setminus\! \{v\} $};

\draw [color=prunelle,decorate,decoration={brace,amplitude=6pt}, line width=1.5pt](8.5,-0.05) -- (7,-0.05);
\draw[color=prunelle](7.8,-0.7) node{$\Bb(u) \!\cap\! T$};


\draw[line width =1.5pt](-1,1)--(-1,-0.2);
\draw (-1,0) node[below=4pt]{$d$} node{$|$};

\tacheCouleurPbas{$u$}{1.5}{7}{olive}
\end{scope}
\end{tikzpicture}
}

\newcommand{\figReInsertE}[1]{
\begin{tikzpicture}[scale=0.7]
\footnotesize
\newcommand{\T}{8};
\newcommand{\retroT}{-8};
\draw(\retroT-1.6,0) node{$\S$};
\draw [->](\retroT-1.2,0)--(\T+0.4,0);
\foreach \x in {1,2,3}  \tacheCouleur{}{1}{\retroT+\x}{orange};
\if#10
\draw[color=orange](-6.5,1) node{$\Ab(u) \!\cap\!E$};
\else
\draw [color=orange,decorate,decoration={brace,amplitude=6pt}, line width=1.5pt](-7.9,0.65) -- (-5.1,0.65) node[midway, above=7pt]{$\Ab(u) \!\cap\!E$};
\fi

\foreach \x in {1,2}  \tacheCouleur{}{1}{-3.5+\x}{oranger};
\tacheCouleur{}{1.5}{0}{oranger};
\if#10
\draw[color=oranger](-1.9,1) node{$A(u) \!\cap\!E$};
\else
\draw [color=oranger,decorate,decoration={brace,amplitude=6pt}, line width=1.5pt](-3.4,0.65) -- (-0.1,0.65) node[midway, above=7pt]{$A(u) \!\cap\!E$};
\fi

\tacheCouleur{}{0.7}{0.7}{magenta};
\tacheCouleur{}{1.3}{2}{magenta};
\if#10
\draw[color=parme](1,1) node{$B(u) \!\cap\!T$};
\else
\draw [color=magenta,decorate,decoration={brace,amplitude=6pt}, line width=1.5pt](0.1,0.65) -- (1.9,0.65) node[midway, above=7pt]{$B(u) \!\cap\!T$};
\fi

\tacheCouleur{}{1}{3}{violet};
\tacheCouleur{}{1.5}{4.5}{violet};
\tacheCouleur{}{1}{5.5}{violet};
\if#10
\draw[color=violet](3.7,1) node{$\Bb(u) \!\cap\!T$};
\else
\draw [color=violet,decorate,decoration={brace,amplitude=6pt}, line width=1.5pt](2.1,0.65) -- (5.4,0.65) node[midway, above=7pt]{$\Bb(u) \!\cap\!T$};
\fi

\draw[line width =1.5pt](0,1)--(0,-0.2);
\draw (0,0) node[below=5pt]{$d$};

\tacheCouleurP{$u$}{1.5}{-3.5}{olive}

\newcommand{\plusBas}{-2.8}
\renewcommand{\z}{-2.8}
\draw(\retroT-1.6,0\plusBas) node{$\S'$};
\draw [->](\retroT-1.2,0\plusBas)--(\T+0.4,0\plusBas);

\foreach \x in {1,2,3}  \tacheCouleur{}{1}{\retroT+\x+1.5}{orange};
\if#10
\draw[color=orange](-5,1\plusBas) node{$\Ab(u) \!\cap\!E$};
\else
\draw [color=orange,decorate,decoration={brace,amplitude=6pt}, line width=1.5pt](-6.4,\z+0.65) -- (-3.6,\z+0.65) node[midway, above=7pt]{$\Ab(u) \!\cap\!E$};
\fi

\foreach \x in {1,2}  \tacheCouleur{}{1}{-3.5+\x}{oranger};
\tacheCouleur{}{1.5}{0}{oranger};
\if#10
\draw[color=oranger](-1.9,1\plusBas) node{$A(u) \!\cap\!E$};
\else
\draw [color=oranger,decorate,decoration={brace,amplitude=6pt}, line width=1.5pt](-3.4,\z+0.65) -- (-0.1,\z+0.65) node[midway, above=7pt]{$A(u) \!\cap\!E$};
\fi

\tacheCouleur{}{0.7}{0.7}{magenta};
\tacheCouleur{}{1.3}{2}{magenta};
\if#10
\draw[color=magenta](1,1\plusBas) node{$B(u) \!\cap\!T$};
\else
\draw [color=magenta,decorate,decoration={brace,amplitude=6pt}, line width=1.5pt](0.1,\z+0.65) -- (1.9,\z+0.65) node[midway, above=7pt]{$B(u) \!\cap\!T$};
\fi

\tacheCouleur{}{1}{\T-3.5}{violet};
\tacheCouleur{}{1.5}{\T-2}{violet};
\tacheCouleur{}{1}{\T-1}{violet};
\if#10
\draw[color=violet](5.2,1\plusBas) node{$\Bb(u) \!\cap\!T$};
\else
\draw [color=violet,decorate,decoration={brace,amplitude=6pt}, line width=1.5pt](3.6,\z+0.65) -- (6.9,\z+0.65) node[midway, above=7pt]{$\Bb(u) \!\cap\!T$};
\fi

\draw[line width =1.5pt](0,1\plusBas)--(0,-0.2\plusBas);
\draw (0,0\plusBas) node[below=4pt]{$d$};

\tacheCouleurP{$u$}{1.5}{3.5}{olive}

\newcommand{\zInterligne}{-1}
\draw[orange,dashed] (-8,0) --(-8,-0.2\plusBas);
\draw[orange,dashed] (-6.5,0) --(-6.5,-0.2\plusBas);
\draw[olive,<->](-8,\zInterligne)--(-6.5,\zInterligne)node[midway, above]{$p_u$};

\draw[violet,dashed] (5.5,0) --(5.5,\zInterligne-0.4);
\draw[violet,dashed] (7,0) --(7,-0.2\plusBas);
\draw[olive,<->](5.5,\zInterligne)--(7,\zInterligne)node[midway, above]{$p_u$};

\draw[oranger,dashed] (-3.5,0) --(-3.5,-0.2\plusBas);
\draw[parme,dashed] (2,0) --(2,-0.2\plusBas);

\end{tikzpicture}
}

\newcommand{\figSolCentered}[1]{
\begin{tikzpicture}[scale=#1]

\foreach \xh/\yh/\l in {
1.5/2.5/A, 2.5/2.5/B, 3.5/2.5/C,
1/2/D, 2/2/E, 3/2/F, 4/2/G,
1.5/1.5/H,  3.5/1.5/J, 
1/1/K, 2/1/L, 3/1/M, 4/1/N
}{
\draw[blue] (\xh,\yh) node{$\bullet$} node[right=1pt]{\scriptsize $\l$};
}
\draw[blue] (2.5,1.5) node{$\bullet$} node[above right =1pt]{\scriptsize $I$};


\tikzstyle{voisin}=[line width=1pt,->,>=latex]
\draw [voisin,orange] (3.5,1.58) -- (3.5,2.47);
\draw [voisin,rose] (3.47,1.57) -- (3.05,2.0);
\draw [voisin,oranger] (3.53,1.57) -- (3.95,2.0);
\draw [voisin,olive] (3.53,1.46) -- (3.95,1.03);
\draw [voisin,turquoise] (3.47,1.46) -- (3.05,1.03);
\draw [voisin,violet] (3.44,1.5) -- (2.56,1.5);
\end{tikzpicture}
}

\newcommand{\figOpeCentered}[1]{
\begin{tikzpicture}[scale=#1]

\foreach \xh/\yh/\l in {
1.5/2.5/A, 2.5/2.5/B, 3.5/2.5/C,
1/2/D, 2/2/E, 3/2/F, 4/2/G,
1.5/1.5/H, 2.5/1.5/I, 3.5/1.5/J,
1/1/K, 2/1/L, 3/1/M, 4/1/N
}{
\draw[blue] (\xh,\yh) node{$\bullet$} node[right=1pt]{\scriptsize $\l$};
}

\foreach \xh/\yh in {1/2,2/2,3/2,1.5/1.5,2.5/1.5,3.5/1.5,1/1,2/1,3/1} {
\draw [oranger, line width=1pt,->,>=latex] (\xh+0.05,\yh+0.075) -- (\xh+0.44,\yh+0.44);
}
\end{tikzpicture}
}




\newcommand{\tblExact}{
\begin{tabular}{c@{}*{8}{p{0.4cm}@{}c@{\enskip}c@{\enskip}c@{}}}
&&\multicolumn{3}{c}{$\FDX$}
&&\multicolumn{3}{c}{$\FI$}
&&\multicolumn{3}{c}{$\FS$} 
&&\multicolumn{3}{c}{$\FIS$}
&&\multicolumn{3}{c}{$\FF$}
&&\multicolumn{3}{c}{$\FFI$}
&&\multicolumn{3}{c}{$\FFS$} 
&&\multicolumn{3}{c}{$\FFIS$}\\[0.1cm]
\cline{3-5}  \cline{7-9}
\cline{11-13}\cline{15-17}
\cline{19-21}\cline{23-25}
\cline{27-29}\cline{31-33}
$n$ 
&& \opt & \time & \nd    && \opt & \time & \nd 
&& \opt & \time & \nd    && \opt & \time & \nd 
&& \opt & \time & \nd    && \opt & \time & \nd 
&& \opt & \time & \nd    && \opt & \time & \nd \\[0.1cm]
\hline \hline
10 
&&10 &  29 & 11 
&&10 &  34 & 10
&&10 &  32 & 7 
&&10 &   0 & 0 
&&10 &  26 & 0
&&10 &  22 & 0
&&10 &   3 & 0 
&&10 &   0 & 0\\
\hline
20 
&&10 &  51 & 162 
&&10 &  63 & 91
&&10 &  63 & 25 
&&10 &  42 & 11 
&&10 &  44 & 0
&&10 &  54 & 0 
&&10 &  41 & 0
&&10 &  10 & 0\\
\hline
50 
&&10 & 311 & 53596 
&&10 &  76 & 2101
&&10 &  90 & 56 
&&10 &  67 & 31 
&&10 &1310 & 24725 
&&10 & 156 & 1293 
&&10 &  15 & 0 
&&10 &  13 & 0\\
\hline
60
&& 5 &2078 & 228193
&&10 & 186 & 8063
&&10 &  74 & 83
&&10 &  58 & 41
&& 0 &   - & -
&& 5 & 439 & 2904 
&&10 &  93 & 66 
&&10 &  15 & 0\\
\hline
80
&& 0 &   - & - 
&& 9 & 815 & 17604
&&10 & 137 & 138
&&10 &  77 & 70
&& - &   - & - 
&& 2 &2823 & 1402 
&&10 & 219 & 322 
&&10 &  79 & 73\\
\hline
100
&&  - &   - & - 
&& 4  &2800 & 23965
&& 10 & 291 & 215 
&& 10 & 109 & 75
&&  - &   - & - 
&&  - &   - & - 
&& 10 & 529 & 542 
&& 10 & 165 & 141\\
\hline
120 
&&  - &  - & - 
&&  - &  - & - 
&&10 & 728 & 269
&&10 & 219 & 122
&&  - &   - & - 
&&  - &   - & - 
&&10 &1578 & 779 
&&10 & 363 & 181\\
\hline
150
&&  - &   - & - 
&&  - &   - & - 
&&  8 &2532 & 410 
&& 10 & 786 & 201 
&&  - &   - & - 
&&  - &   - & - 
&&  2 &3172 & 660 
&& 10 &1011 & 481
\\
\hline
180
&&  - &   - & - 
&&  - &   - & - 
&&  1 &3514 & 285 
&&  6 &2460 & 194 
&&  - &   - & - 
&&  - &   - & - 
&&  - &   - & - 
&&  5 &1537 & 284 \\
\hline
200
&&  - &   - & - 
&&  - &   - & - 
&&  - &   - & - 
&&  1 &1929 & 127 
&&  - &   - & - 
&&  - &   - & - 
&&  - &   - & - 
&&  4 &2524 & 710 \\
\end{tabular}
}

\newcommand{\tblLB}{
\begin{tabular}{c*{9}{p{0.15cm}@{}c@{}p{0.15cm}@{}c}}
&&\multicolumn{3}{c}{$\FDXLP$}
&&\multicolumn{3}{c}{$\FDXRN$}
&&\multicolumn{3}{c}{$\FIRN$}
&&\multicolumn{3}{c}{$\FSRN$} 
&&\multicolumn{3}{c}{$\FISRN$}
&&\multicolumn{3}{c}{$\FFRN$}
&&\multicolumn{3}{c}{$\FFIRN$}
&&\multicolumn{3}{c}{$\FFSRN$} 
&&\multicolumn{3}{c}{$\FFISRN$}\\[0.1cm]
\cline{3-5}   \cline{7-9}
\cline{11-13} \cline{15-17}
\cline{19-21} \cline{23-25}
\cline{27-29} \cline{31-33}
\cline{35-37}
$n$ 
&& \LB && \time && \LB && \time 
&& \LB && \time && \LB && \time 
&& \LB && \time && \LB && \time 
&& \LB && \time && \LB && \time 
&& \LB && \time \\[0.1cm]
\hline \hline
10 
&& 41\% && 0
&& 41\% && 0
&& 33\% && 0
&& 41\% && 0 
&&  0\% && 0
&&  7\% && 1
&&  5\% && 1
&&  0\% && 1
&&  0\% && 0
\\
\hline
20 
&& 68\% && 0
&& 68\% && 0
&& 66\% && 0
&& 68\% && 0 
&& 12\% && 1
&& 28\% && 2
&& 28\% && 2
&&  6\% && 1 
&&  2\% && 0
\\
\hline
50 
&& 86\% && 0
&& 86\% && 1
&& 86\% && 1
&& 86\% && 6 
&& 28\% && 6
&& 42\% && 27
&& 41\% && 31
&& 17\% && 5 
&& 11\% && 3
\\
\hline
60 
&& 89\% && 0
&& 89\% && 1
&& 89\% && 1
&& 89\% && 7 
&& 36\% && 7
&& 41\% && 91
&& 41\% && 95
&& 22\% && 9
&& 16\% && 5
\\
\hline
80
&& 92\% && 1
&& 92\% && 1
&& 92\% && 1
&& 92\% && 11 
&& 34\% && 8
&& 43\% && 345
&& 43\% && 359
&& 21\% && 28
&& 15\% && 10
\\
\hline
100 
&& 93\% && 2 
&& 93\% && 2
&& 93\% && 2
&& 93\% && 19 
&& 35\% && 14
&& 45\% && 1091 
&& 44\% && 1152
&& 21\% && 62
&& 14\% && 25 
\\
\hline
120 
&& 94\% && 3
&& 94\% && 4
&& 94\% && 11
&& 94\% && 31 
&& 38\% && 15
&& 46\% && 3189
&& 46\% && 3192
&& 22\% && 133
&& 16\% && 52
\\
\hline
150
&& 96\% && 6
&& 96\% && 13
&& 96\% && 15
&& 96\% && 60 
&& 34\% && 29
&&    - && - 
&&    - && - 
&& 22\% && 352
&& 15\% && 130  
\\
\hline
180
&& 96\% && 12
&& 96\% && 19
&& 96\% && 23
&& 96\% && 98 
&& 34\% && 49
&&    - && - 
&&    - && - 
&& 22\% && 766
&& 15\% && 274  
\\
\hline
200
&& 97\% && 19
&& 97\% && 25
&& 97\% && 31
&& 97\% && 126 
&& 39\% && 72
&&    - && - 
&&    - && - 
&& 22\% && 1204 
&& 15\% && 418  
\\
\hline
500
&& {99\%}* && 722
&& {99\%}* && 698
&& {99\%}* && 742
&& {99\%}* && 2820
&& {36\%}* && 1870
&&       - && - 
&&       - && - 
&&       - && - 
&&       - && - 
\end{tabular}
}

\newcommand{\tblUBvUn}{
\begin{tabular}{c *{3}{p{0.8cm}@{}c@{}p{0.4cm}@{}c@{ }} }
&& $\HBF$ && $\HBFp$  
&& $\HR$  && $\HRp$
&& $\HRR$  && $\HRRp$\\
\cline{3-3}   \cline{5-5} 
\cline{7-7}   \cline{9-9}
\cline{11-11} \cline{13-13}
$n$ 
&& \UB && \UB 
&& \UB && \UB 
&& \UB && \UB  \\[0.1cm]
\hline \hline
10 
&& 2.04\% && 0.00\%  
&&  170\% && 0.00\%
&& 0.00\% && 0.00\% \\
\hline
20 
&& 0.95\% && 0.00\% 
&&  196\% && 0.00\% 
&& 1.33\% && 0.00\% \\

\hline
50 
&& 0.35\% && 0.02\%  
&&  203\% && 0.00\%  
&&13.83\% && 0.00\% \\

\hline
60 
&& 0.26\% && 0.01\%  
&&  170\% && 0.01\%  
&&16.80\% && 0.01\% \\

\hline
80 
&& 0.22\% && 0.01\%  
&&  172\% && 0.00\%  
&&16.36\% && 0.00\% \\

\hline
100 
&& 0.18\% && 0.00\% 
&&  174\% && 0.00\%   
&&15.72\% && 0.00\% \\

\hline
120 
&& 0.10\% && 0.00\%  
&&  170\% && 0.00\%  
&&15.77\% && 0.00\% \\

\hline
150 
&& 0.10\% && 0.00\% 
&&  171\% && 0.00\%  
&&15.27\% && 0.00\% \\

\hline
180 
&& 0.10\% && 0.00\%  
&&  171\% && 0.00\%  
&&16.09\% && 0.00\% \\

\hline
200
&& 0.10\% && 0.01\% 
&&  171\% && 0.01\%  
&&16.28\% && 0.00\% \\

\end{tabular}
}

\newcommand{\tblCcl}{
\begin{tabular}{c
*{2}{p{0.6cm}@{}c@{}p{0.3cm}@{}c@{}p{0.3cm}@{}c@{ }}
p{0.6cm}@{}p{0.3cm}c@{}p{0.3cm}@{}c@{ }
p{0.6cm}@{}p{0.15cm}c@{}p{0.3cm}@{}c@{ }
}
to obtain:
&&\multicolumn{5}{c}{an upper bound}
&&\multicolumn{5}{c}{a lower bound}
&&\multicolumn{4}{c}{a 5\%-approximation}
&&\multicolumn{4}{c}{an exact solution}\\
\cline{3-7}\cline{9-13}\cline{15-18}\cline{20-23}\\[-0.5cm]
use:
&&\multicolumn{5}{c}{\FA}
&&\multicolumn{5}{c}{\FB}
&&\multicolumn{4}{c}{\FC}
&&\multicolumn{4}{c}{\FD}\\[0.1cm]
\cline{3-7}\cline{9-13}\cline{15-18}\cline{20-23}\\[-0.5cm]
$n$ 
&& \LB && \UB && time 
&& \LB && \UB && time
&&& \time && \nd
&&& \time && \nd \\[0.1cm]
\hline \hline
50 
&& 86\% && 0.00\% && <1 
&& 11\% && 0.00\% &&  3
&&&   8 && 0       
&&&   4 && 34 \\
\hline
100 
&& 93\% && 0.00\% && 2
&& 14\% && 0.00\% && 25 
&&& 160 &&  114  
&&& 165 &&  141 \\
\hline
200 
&& 97\% && 0.01\% && 20
&& 15\% && 0.01\% &&418 
&&& 7420 && 928    
&&& 8317 && 1474  \\  
\hline
500 
&& - && -(99\%) && 778
&& - && - && -
&&& - && -
&&& - && -
\end{tabular}
}

\newtheorem[M,cut=false]{df}{Definition}
\newtheorem[M,cut=false]{pte}[df]{Property}
\newtheorem[M,cut=false]{cor}[df]{Corollary}
\newtheorem[M,cut=false]{lm}[df]{Lemma}
\newtheorem[M,cut=false]{thm}[df]{Theorem}
\newtheorem[M,cut=false]{rpl}[df]{Remind}
\newtheorem[M,cut=false]{nota}[df]{Notations}
\newtheorem[M,cut=false]{conj}[df]{Conjecture}
\newtheorem[M,cut=false]{ce}{Counter example}
\renewcommand {\myproofname}{Proof}
\renewcommand {\claimname}{Claim}

\titleformat{\subsubsection}{\it}{$\bullet$}{0.15cm}{}
\titlespacing{\subsubsection}{+0pc}{3.5ex plus .1ex minus .2ex}{0.8ex minus .1ex}

\begin{frontmatter}

\title{ 
Dominance inequalities for scheduling\\
around an unrestrictive common due date
}

\author[SU]{Anne-Elisabeth Falq\corref{corresp}}
\cortext[corresp]{Corresponding author}
\ead{anne-elisabeth.falq@lip6.fr}
\author[SU]{Pierre Fouilhoux\corref{corresp}}
\ead{pierre.fouilhoux@lip6.fr}
\author[CNAM]{Safia Kedad-Sidhoum\corref{corresp}}
\ead{safia.kedad_sidhoum@cnam.fr}

\address[SU]{Sorbonne Universit\'e, CNRS, LIP6,
4 Place Jussieu, 75005 Paris, France}
\address[CNAM]{ CNAM, CEDRIC,
292 rue Saint Martin, 75141 Paris Cedex 03, France}

\begin{abstract}
\noindent
The problem considered in this work
consists in scheduling 
a set of tasks 
on a single machine,
around an unrestrictive common due date 
to minimize the 
weighted sum of earliness and tardiness.
This problem can be formulated as 
a compact mixed integer program (MIP).
In this article,
we focus on neighborhood-based dominance properties,
where the neighborhood is associated
to insert and swap operations.
We derive from these properties a local search procedure
providing a very good heuristic solution.
The main contribution of this work
stands in an exact solving context:
we derive constraints eliminating 
the non locally optimal solutions
with respect to the insert and swap operations.
We propose linear inequalities
translating these constraints
to strengthen the MIP compact formulation. 
These inequalities, called dominance inequalities,
are different from standard reinforcement inequalities.
We provide a numerical analysis which shows
that adding these inequalities significantly
reduces the computation time required
for solving the scheduling problem 
using a standard solver.
\end{abstract}


\end{frontmatter}


\section{Introduction}
The scheduling problem studied in this work
falls within the just-in-time scheduling field.
In this framework, 
each task has a due date,
and any deviation from this due date is penalized.
From one hand,
the tardiness is penalized to model the customer dissatisfaction,
on the other hand,
the earliness is penalized to model the 
induced storage costs .
The reader can refer to the seminal surveys 
of~\cite{Baker_Scudder_90} and~\cite{survey_17}
for the early results in this field.\\

\AF{le problème}
We consider a set $J$ of $n$ tasks  
with fixed processing times $(p_j)_{j\in J} \iin \Rp^J$,
to be non-preemptively processed
on a single machine.
These tasks share a common \textit{due date} $d$
for which tasks should be preferably completed.
In this work,
we assume that the due date is \textit{unrestrictive},
\ie  $d \!\geqslant\! \sum p_j$.

In a just-in-time framework,
tasks completing before or after $d$ will therefore
incur penalties according to unit earliness (\resp tardiness) penalties
$(\a_j)_{j\in J} \iin \Rp^J$
(\resp  $(\b_j)_{j\in J} \iin \Rp^J$).
A schedule is defined by the task completion times
denoted by $(C_j)_{j\in J}$.
Using $[x]^+$ to denote the positive part
 of $x \!\in\! \mathbb{R}$,
the \textit{earliness} (\resp \textit{tardiness})
of a task $j\!\in\! J$ 
is given by $[d\!-\!C_j]^+$ (resp. $[C_j\!-\!d]^+$).
Given parameters $d$, $p$, $\a$ and $\b$,
the Unrestrictive Common Due Date Problem (\lepb)
aims at finding  a schedule minimizing 
the total \textit{penalty}:
$$\sum\limits_{j\in J}
\alpha_j\,[d \!-\! C_j]^+ \!+ \beta_j \,[C_j \!-\! d]^+$$
This criterion is non-regular,
\ie  is not a non-increasing function
of completion time $C_j$ for any $j\iin J$.
Moreover,
the penalty function
\modifRtrois{is not  a linear} function
 of the completion times.\\



\AF{résultats de complexité}
If all task penalties are equal,
\ie $\alpha_j\!=\!\beta_j$ for any task $j\iin J$, 
the \lepb is solvable in polynomial time~\citep{Kanet}.
If task penalties are symmetric,
\ie  $\alpha_j \!=\! \beta_j$ for all $j\!\in\! J$,
the problem is NP-hard~\citep{Hall_et_Posner}.
Therefore,
the problem that we consider
with arbitrary $\a$ and $\b$ coefficients 
is also NP-hard.


\AF{état de l'art PL}
A wide range of variables can be used 
to formulate a single machine scheduling problem 
as a {Mixed Integer Program (MIP)}:
completion time variables,
time-indexed variables,
linear ordering variables,
positional date and 
assignment variables~\citep{Queyranne_Schulz_94}.
However, 
only few works based on linear programming are proposed
for the \lepb.
\modifRdeux{
\cite{benchmark} propose a compact MIP formulation
based on disjunctive variables,
which is only given to compare 
the proposed heuristic method to an exact method for small instances.
\cite{Van_den_Akker_HVDV_2002} propose a formulation
based on an exponential number of binary variables
using column generation and Lagrangian relaxation.
}
In~\cite{nous19},
we propose a MIP
based on natural variables,
similar to completion time variables,
together with a compact MIP based on partition variables.\\

\AF{état de l'art ordo(issu de DAM)}
Furthermore,
the \lepb have been solved through 
several other dedicated exact methods
like branch-and-bound algorithms 
(\textit{e.g.}~\cite{Sourd_09})
and dynamic programming methods 
(\textit{e.g.}~\cite{Hall_et_Posner}, \cite{HVDV}, \cite{Tanaka_Araki_13}).
Moreover,
a benchmark is provided in~\cite{benchmark},
together with a heuristic method.
An important remark is that 
the efficiency of the latter approaches
comes from the exploitation of dominance properties.
In particular,
the dedicated \BB proposed in~\cite{Sourd_09}
for an exact solving of \lepb, 
solves up to 1000-tasks instances 
of the benchmark provided in~\cite{benchmark}.\\

\AF{dominances utilisées en encodage et en inégalités}
Among the dominance properties,
we can distinguish the structural ones,
which allow to restrict the solution set 
to the set of the schedules having a specific structure.
These properties are already taken into account
in the compact formulation 
that we propose in~\cite{nous19}.
Indeed, thanks to their structure,
the dominant schedules are encoded by
only $n$ binary variables
resulting in a formulation 
whose size  \modifRtrois{does not depend} on the time horizon,
in contrast with the time-indexed formulations.
In the present work,
we focus on another type of dominance property
called neighborhood based dominance properties.
We propose to model them
by linear inequalities.\\ 

\AF{inégalités pas comme les autres = contributions}
One important contribution of this article
is to propose inequalities strengthening the
linear formulation in a new way.
They are different from standard reinforcement 
and symmetry-breaking inequalities.
\modifRtrois{The standard reinforcement inequalities}
cut fractional points
to improve the lower bound 
obtained from the linear relaxation,
\modifRtrois{the symmetry-breaking inequalities cut}
integer points,
- which may be optimal-
to reduce the search space.
In contrast, 
Dominance inequalities aims at cutting
sub-optimal integer points. \\

\AF{Organisation du papier}
This article is organized as follows.
In Section~\ref{sec_dom},
we recall structural dominance properties
leading to reformulate the \lepb
into a partition problem; 
then we adapt some neighborhood based dominance properties
used on schedules for partitions.
Section~\ref{sec_dom_PL} translates in a linear way 
the constraints proposed in Section~\ref{sec_dom},
which leads to a new compact linear formulation,
enriched with dominance inequalities.
Section~\ref{sec_expe} presents some experimental results
to show the contribution of considering dominance in
an exact solving process, as in a heuristic approach.

\section{Dominance properties}
\label{sec_dom}
A subset of solutions is said \textit{dominant} 
if it contains at least one optimal solution,
and \textit{strictly dominant} 
if it contains all the optimal solutions.
For brevity,
a schedule will be said dominant (\resp strictly dominant)
if it belongs to a dominant (\resp strictly dominant) set.
In this article,
we will only use dominance properties about solutions,
even if there also exist
dominance properties about instances or problems~\citep{Jouglet_Carlier_11}.

\subsection{Structural dominance properties}
\AF{définitions}
In order to describe dominant schedules,
we first provide some useful definitions.
\modifRun{
A task $j\!\in\! J$ is said \textit{early} 
(\resp \textit{tardy}) 
if it completes at time $C_j\!\leqslant\! d$
( \resp  $C_j\!>\! d$).
Among the early tasks, 
the one completing at $d$ (if it exists),
is called the \textit{on-time} task.
}
The \textit{early-tardy partition} of a schedule 
is the pair $(E,T)$ where
$E$ (\resp $T$) is the early (\resp tardy) task subset.
Moreover,
we will use \textit{$\a$-ratio} (\resp \textit{$\b$-ratio})
to designate $\a_j/p_j$ (\resp $\b_j/p_j$).
We define a \textit{block} 
as a feasible schedule without idle time between task execution, 
a \hbox{\textit{$d$-schedule}} 
as a feasible schedule with an on-time task,
and a \textit{\dbloc} 
as a $d$-schedule which is also a block.
A schedule is said \textit{\Vshaped}
if early tasks are scheduled
in non-decreasing order of their $\a$-ratios 
and the tardy ones 
in non-increasing order of their $\b$-ratios.\\ 

\AF{dominances}
The following lemma gives dominance properties 
already known for the unrestrictive common due date problem
with symmetric penalties~\citep{Hall_et_Posner}.
These results have been extended to asymmetric penalties
in~\cite{nous19},
using the same task shifting and exchange
proof arguments.

\begin{lm}[\cite{nous19}]
The set of \Vshaped schedules is strictly dominant 
for the \lepb.\\[0.05cm]
The set of \dblocs is dominant for the \lepb. 
\modifRun{
Moreover, if unit earliness and tardiness penalties are positive,
\ie $(\a,\b)\iin {\Rpe}^J \!\!\times\! {\Rpe}^J$,
then the set of blocks is strictly dominant.
}
\label{dom_unres}
\end{lm}

\AF{hypothèses sur les entrées}
Thanks to the \dbloc dominance,
we can make some assumptions on parameters $p,\a,\b$
without  loss of generality.
A task having a zero processing time
can be inserted between two tasks of a schedule
without impacting other tasks.
In particular in a \dbloc,
such a task can be inserted at the due date,
which incurs no penalty variation.
Hence,
we assume that $\forall j\iin J,\, p_j\iin\Rpe$.\\
Since the due date $d$ is large enough,
a task having a zero earliness (\resp tardiness) penalty
can be added at the beginning (\resp at the end)
of a \dbloc without incurring any cost variation.
Hence,
we assume that $\forall j\iin J, (\a_j,\b_j)\iin(\Rpe)^2$.\\

\AF{se ramener à un problème de partition, classe d'équivalence, introduire $f$}
Since they form a dominant set,
we only consider \Vdblocs in the following.
Note that in such schedules,
tardy tasks are completely processed after $d$,
since there is no straddling task,
\ie there is no task starting before $d$ 
and completing after $d$.
Therefore, the early (\resp tardy) task set 
can be referred to the left (\resp right) side of $d$.
\\

In general,
a \Vdbloc cannot be encoded by its early-tardy partition. 
Indeed, 
if two early (\resp tardy) tasks
have the same $\a$-ratio (\resp $\b$-ratio),
one cannot determine which one is processed first,
they can be sequenced arbitrarily.
However,
swapping two such tasks in a \Vdbloc
results in another \Vdbloc having the same penalty.
Furthermore,
all the \Vdbloc having the same early-tardy partition
have the same penalty.
Based on this remark,
two \Vshaped \hbox{\dblocs} having the same early-tardy partition
will be said \textit{equivalent}.
We denote by $\sim$ this relation.\\

We define an \textit{ordered bi-partition}
of a set $A$ as a couple $(A^1,A^2)$
where $\{A^1,A^2\}$ is a partition of $A$.
This is not only a partition into two subsets
since the two subsets are not symmetric,
\ie $(A^1,A^2) \!\neq\! (A^2,A^1)$.
Note that the early-tardy partition of a schedule 
is an ordered bi-partitions.
More precisely,
there is a one-to-one correspondence
between equivalence classes for $\sim$ 
and ordered {bi-partitions} $(E,T)$ of $J$
where $E\!\neq\! \emptyset$.
Indeed,
the set of the early tasks of a \dbloc
cannot be empty since it contains at least the on-time task.
Let $\Part$ denote the set of such $(E,T)$,
and $\PartC$ denote the set of any ordered {bi-partition} of $J$.
In the sequel, 
we will only say \textit{partition} 
to refer to an ordered {bi-partition}
or to an early-tardy partition,
when there is no ambiguity.\\

For any partition $(E,T) \iin \Part$,
let $f(E,T)$ denote the penalty of the equivalence class $(E,T)$,
that is the penalty of any \Vdbloc of partition $(E,T)$. 
For sake of consistency,
we extend the definition of $f$ to $\PartC$
by setting $f(\emptyset,J)$ as the penalty of 
any \Vshaped block starting at time $d$. 
From now on, 
our aim is to find 
a partition of $J$
minimizing $f$, since the \lepb 
can be formulated as follows.
$$(\FET):\enskip{\min_{(E,T)\,\in\,\Part} f(E,T)}$$

\subsection{Neighborhood based dominance properties}


\noindent\AF{idée dominance des solutions non dominées dans leur voisinage}
Let us recall some definitions used in local search context~\citep{Aarts_Lenstra_03}.
A \textit{neighborhood function} $N$
is a function which associates to any solution $\S$
a subset of solutions $N(\S)$,
called the \textit{neighborhood} of $\S$.
A solution of $N(\S)$
is called a \textit{neighbor} of $\S$.
Moreover,
a solution $\S$ is \textit{locally optimal}
with respect to minimizing function $\phi$
if  $\phi(\S) \!\leqslant\! \phi(\S')$,
for any neighbor $\S'\iin N(\S)$.
\modifRtrois{
If, on the contrary, there exists $\S'\iin N(\S)$
such that $\phi(\S') \!<\! \phi(\S)$,
$\S$ is \textit{dominated} (by $\S'$).
}\\

Given a neighborhood function,
the set of locally optimal solutions 
always contains all optimal solutions,
and is therefore a strictly dominant set.
This statement can be seen as a generic dominance property
\modifRtrois{denoted $\mathcal{G}$}.
This kind of dominance property
is commonly used in local search.
Indeed, a step of a local search procedure
consists in enumerating some of the neighbors of a given solution $\S$,
computing their values 
and, if a better solution is found,
then moving to the best solution found
at the current iteration,
which is equivalent to discard 
$\S$ \modifRtrois{since} it is dominated.\\

\subsubsection{Operation-based neighborhoods}
\label{opposition_pdv}
We call \textit{operation}
any (eventually partial) function,
which maps a solution to another solution.
In this work, 
we will consider neighborhood functions
based on a set of operations.
The neighborhood of a solution $\S$ is then
the set of the solutions obtained 
by applying to $\S$ any operation defined on $\S$.
This kind of neighborhood functions allows to 
use the generic dominance property \modifRtrois{$\mathcal{G}$}
in a different way.
Instead of considering sequentially
the neighborhood of each solution,
we will consider sequentially each operation.
For each one, 
any solution is compared to its neighbor
obtained using this operation.
This differentiates the solution-centered 
and the operation-centered point of view,
\modifRtrois{
as developed in the next section/point.}
\\

\subsubsection{The solution-centered and the operation-centered points of view}
\label{opposition_pdv}

Figures~\ref{fig_sol} and~\ref{fig_ope} illustrate
these two points of view
on the same set of solutions $\{A,B,\dots,N\}$
represented by the blue points.
We consider a given set of operations
and the associated neighborhood function.
An arrow is set from a solution $X$ to a solution $Y$
if $X$ is compared to $Y$ 
in order to determine whether $X$ is dominated by $Y$.
\\
In Figure~\ref{fig_sol},
we focus on one solution, $J$,
which is compared to all the solutions obtained
by applying to $J$ an operation defined on $J$.
Since $J$ is compared to all its neighborhood,
\ie $\{C,F,G,I,M,N\}$,
one can determine if $J$ is locally optimal.
\\
In Figure~\ref{fig_ope},
we focus on one operation.
All the solutions where this operation can be applied
are compared with the obtained neighbors.
Solutions $A,B,C,G$ and $N$ are not 
compared to others solutions,
since the considered operation cannot be applied
on these solutions.
Conversely,
since solutions $D,E,F,H,I,J,K,L$ and $M$
are compared to one of their neighbors,
they might be discarded.
For example if $E$ is better than $H$,
$H$ can be discarded, no matter whether
$H$ is better than $K$.
However,
we cannot say 
that the non-discarded solutions are locally optimal,
since only one neighbor is taken into account.
To this end,
all the operations must be considered.


\begin{figure}[h!]
\centering
\captionsetup{justification=centering}
\begin{minipage}{0.4\textwidth}
\centering
\figSolCentered{1.6}
\caption{Illustration of the \textit{solution-centered} point of view}
\label{fig_sol}
\end{minipage}
\hspace*{1cm}
\begin{minipage}{0.4\textwidth}
\centering
\figOpeCentered{1.6}
\caption{Illustration of the \textit{operation-centered} point of view}
\label{fig_ope}
\end{minipage}
\end{figure}

\modifRtrois{
Let us now present the main ideas of a generic method
to handle operation-based dominance property.
First, find operations inducing an objective function variation that can be explicitly expressed.
Then derive, for each operation, 
a constraint ensuring that any solution
is not dominated by applying this operation.
Using all the obtained constraints,
only the dominant set of property \modifRtrois{$\mathcal{G}$} subsists
as all the non-locally optimal solutions have been removed
(where local is understood with respect to
the neighborhood defined by the operations). 
\\
In Section~\ref{subsec_ope}, we introduce
two families of operations over the partitions,
and follow this method, resulting in constraints~\eqref{ctr_insert_av},~\eqref{ctr_insert_retard} 
and~\eqref{ctr_swap} translating Property~\ref{pte_dom}.
}

\subsection{Operations for early-tardy partitions}
\label{subsec_ope}

\AF{deux opérations : définitions + ptés}
In the case of problems where the solutions are partitions,
two operations are commonly 
considered to define a neighborhood:
the \textit{insertion},
which consists in moving 
\modifRtrois{two elements, one in each subset}
and the \textit{swap}
which consists in swapping 
two elements of each subset.
Let us define solutions which are locally optimal in
the neighborhoods associated \modifRtrois{with} these operations. \\

\begin{df}
Let $(E,T) \iin \Part$.\\[-0.2cm]
(i) $(E,T)$ is an \textit{insert local optimum}
if 
$\left\lbrace \begin{tabular}{@{}c@{ }l}
$\forall v\iin  T,$&
$f\big(E,\,T\big) \leqslant f\big(E\!\cup\!\{v\},\,T\!\setminus\!\{v\}\big),$\\[0.2cm]
$\forall u\iin  E,$&
$f\big(E,\,T\big) \leqslant f\big(E\!\setminus\!\{u\},\,T\!\cup\!\{u\}\big)$.
\end{tabular}\right.$
\\[0.2cm]
(ii) $(E,T)$ is a \textit{swap local optimum}
if \\
\hspace*{0.6cm}
\hbox{$\,\forall u\iin E ,\, \forall v\iin  T,\, 
f\big(E,\,T\big) \!\leqslant\! f\big(E\!\setminus\!\{u\}\!\cup\!\{v\},\,T\!\cup\!\{u\}\!\setminus\!\{v\}\big)$}.
\end{df}



\begin{pte}
The set of insert locally optimal partitions,
as well as the set of swap locally optimal partitions,
is strictly dominant when minimizing $f$ over $\Part$.
\label{pte_dom}
\end{pte}

In terms of scheduling,
the insert operation consists in 
removing a  task $j \iin J$ 
from the  early (\resp tardy) side
and inserting $j$ on the tardy (\resp early) side,
as early (\resp as tardy) as possible 
according to its $\b$-ratio (\resp $\a$-ratio).
The tasks scheduled before (\resp after) $j$ 
are right-shifted (\resp left-shifted) 
by $p_j$ time units.
The swap operation consists in  
sequentially inserting an early task on the tardy side
and inserting another tardy task on the early side,
or vice-versa,
as described above.
Let us denote by $\text{insert}_u(\S)$ 
(\resp $\text{swap}_{u,v}(\S)$)
the schedule obtained from a schedule $\S$
by the insert (\resp swap) operation 
on task $u$ (\resp on tasks $u,v$).\\

Since the insert and swap operations are 
fundamentally defined on partitions
(\ie over equivalence classes of \Vdblocs),
these operations preserve the equivalence.
Therefore we have 
$\forall u\iin J,\,
\S \!\sim\! \S' \Rightarrow \text{insert}_u(\S) \!\sim\! \text{insert}_u(\S')$
and
$\forall (u,v)\iin J^2,\,
\S \!\sim\! \S' \Rightarrow \text{swap}_{u,v}(\S) \!\sim\! \text{swap}_{u,v}(\S')$,
assuming that $u$ and $v$ are 
not on the same side in these schedules.
\\



\AF{intro du représentant favori, notations A B,structure des \Vdblocs}
In order to provide an expression of
the penalty variation induced by such an operation,
\modifRun{it is convenient} to choose a specific \Vdbloc as representative
of an \modifRtrois{equivalence class.}
More precisely,
the chosen representative will depend
on the inserted task (\resp on the swapped tasks).
Let us introduce some notations, 
related to a given task $u\iin J$.
To describe the early side of a \Vdbloc
with regard to $u$, 
the set of remaining tasks $J\!\setminus\!\{u\}$
can be split
according to their $\a$-ratio
into two subsets
$A(u)$ and $\Ab(u)$ 
defined as follows.
$$A(u)  \= \left\{i\iin J\,\middle|\, \frac{\a_i}{p_i} \!>\! \frac{\a_u}{p_u}\right\}
\enskip \text{ and } \enskip 
\Ab(u)\= \left\{i\iin J \!\setminus\!\{u\}\,\middle|\, \frac{\a_i}{p_i} \!\leqslant\! \frac{\a_u}{p_u}\right\}$$ 
Similarly,
we introduce the two following subsets
to describe the tardy side.
$$B(u)  \= \left\{i\iin J\,\middle|\, \frac{\b_i}{p_i} \!>\! \frac{\b_u}{p_u}\right\}
\enskip \text{ and } \enskip 
\Bb(u)\= \left\{i\iin J \!\setminus\!\{u\}\,\middle|\, \frac{\b_i}{p_i} \!\leqslant\! \frac{\b_u}{p_u}\right\}$$

\noindent
Note that if $u$ is early in a \Vdbloc,
early tasks belonging to $A(u)$ are 
necessarily scheduled after $u$,
because of their $\a$-ratio.
To illustrate,
we can observe in Figure~\ref{fig_insert_av}
that the tasks of $\Au$
are scheduled after $u$ in the schedule $\S$. 
Conversely, an early task of $\Ab(u)$
is not necessarily scheduled before $u$,
when its $\a$-ratio is the same as $u$.
However,
we will consider a representative
where $u$ is placed after all the early tasks of $\Ab(u)$,
that is as tardy as possible according to its $\a$-ratio.
Similarly, in case where $u$ is tardy,
we will consider a representative
where $u$ is scheduled before all tardy tasks of $\Bb(u)$.
Such a representative will be called
a $u$-\textit{canonical} \Vdbloc.
The schedule $\S$ (\resp $\S'$) in Figure~\ref{fig_insert_av}
represents the shape of a $u$-{canonical} \Vdbloc
when $u$ is early (\resp tardy).

%

\subsubsection{Penalty variation induced by an insert operation}
\noindent
Given $u\iin J$,
let $(E,T)$ be a partition such that $u\iin E$.
We aim to express the variation of $f$ 
induced by the insertion of $u$ in T, 
\ie between $(E,T)$ and $(E\!\setminus\!\{u\}, T\!\cup\!\{u\})$.
To this end,
we consider a $u$-canonical representative $\S$ of $(E,T)$,
and the \Vdbloc $\S'$ obtained from $\S$ 
by inserting $u$ in $T$, 
as early as possible,
that is just after tardy tasks of $B(u)$.
Note that
$\S'$ is thus a $u$-canonical representative of 
$\big(E\!\setminus\!\{u\},T\!\cup\!\{u\}\big)$.
Let $(e,t)$ (\resp $(e',t')$) denote
the earliness and tardiness vector
of tasks in $\S$ (\resp $\S'$).\\

\begin{figure}[h]
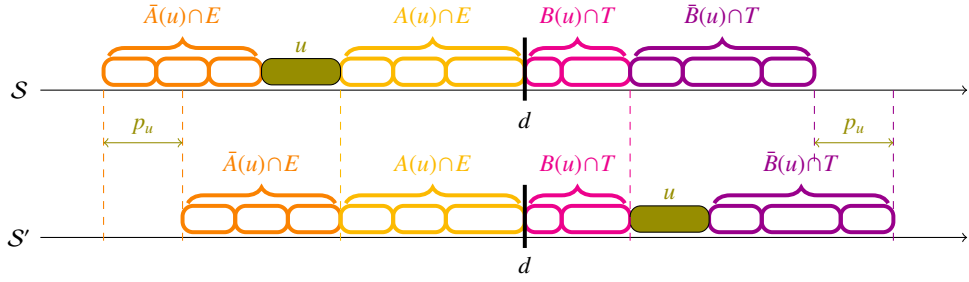

\figReInsertE{1}\\[-0.6cm]
\caption{Illustration of the insert operation of an early task $u$ on the tardy side}
\label{fig_insert_av}
\end{figure}

As we can observe in Figure~\ref{fig_insert_av},
early tasks of $A(u)$
and tardy tasks of $B(u)$
are identically scheduled in $\S$ and $\S'$.
The penalty variation induced by the insertion of $u$ 
is then only due to the move of $u$
and 
to the right-shifting of early tasks of $\Ab(u)$
and tardy tasks of $\Bb(u)$.

Each task $j\iin\Abu$ of $\S$
is postponed $p_u$ time units later in $\S'$,
while staying early.
We then have $e'_j \= e_j - p_u$.
The earliness penalty of task $j$ in $\S'$
is therefore $\a_j\, e'_j =\a_j e_j - \a_j p_u$,
which represents a reduction of $\a_j\,p_u$
compared to its earliness penalty in $\S$.
Summing up over $\Abu$,
we obtain a reduction of the earliness penalties 
of $p_u\, \a\big(\Abu\big)$,
where $x(I) \= \sum_{i\in I} x_i$ 
for any  $x\iin \mathbb{R}^n$
and any $I \inclus [1..n]$,
Similarly,
the right-shifting of tasks in $\Bbu$
induces an increase of the tardiness penalties 
of $p_u\, \b\big(\Bbu\big)$,
since for each $j\iin\Bbu$
we have $\b_j\, t'_j = \b_j\, (t_j \+ p_u)$.\\[-0.2cm]

Moreover,
since $e_u \= p\big(\Au\big)$,
removing $u$ from the early side 
induces a reduction of its earliness penalty 
of $\a_u \,p\big(\Au\big)$.
Similarly,
introducing $u$ on the tardy side
induces an increase of its tardiness penalty
of $\b_u\,t'_u = \b_u\, \Big(p\big(\Bu\big) \+ p_u\Big)$
since $t_u \=0$.\\

\noindent
Finally, 
the penalty variation between $\S$ and $\S'$,
is given by the following expression.\\[-0.4cm]
$$
\Du = -\a_u \, p\big(\Au\big)
+\b_u \, \Big(p\big(\Bu\big) \+ p_u \Big)
+ p_u\, \Big( \b\big(\Bbu\big)- \a\big(\Abu\big) \Big)
$$

\noindent
Since $\S$ and $\S'$  
are representative of $(E,T)$ and 
$(E\!\setminus\!\{u\}, T\!\cup\!\{u\})$ respectively,
$\Du$ is also the variation of $f$ 
induced by the insertion of $u$ in $T$.
Property~\ref{pte_insert}.\textsl{(i)} follows.\\

The insert operation of the tardy task $u$ on the early side 
applied on schedule $\S'$ 
provides schedule $\S$.
Note that
this statement requires that $\S$ is $u$-canonical.
This observation allows to establish
that the penalty variation induced
by inserting $u$ on the early side in $\S'$
is simply $-\Du$,
and results in Property~\ref{pte_insert}.\textsl{(ii)}.

\begin{pte}
Let $(E,T)$  be a partition.\\
(i) For any $u\iin E$,
$f(E\!\setminus\!\{u\}, T\!\cup\!\{u\}) = f(E,T) + \Du$.\\[0.1cm]
(ii) For any $v\iin T$,
$f(E\!\cup\!\{v\}, T\!\setminus\!\{v\}) = f(E,T) - \Dv$.
\label{pte_insert}
\end{pte}

\noindent
Let us introduce, 
for a given task $u\iin J$,
the two following constraints.\\[-0.4cm]
\begin{equation}
u\iin E \Rightarrow\Du \!\geqslant\! 0 
\tag{$I_u$} \label{ctr_insert_av}
\end{equation}
\vspace*{-0.6cm}
\begin{equation}
u\iin T \Rightarrow \Du \!\leqslant\! 0 
\tag{$I'_u$} \label{ctr_insert_retard}
\end{equation}
\noindent
Thanks to Property~\ref{pte_insert},
\eqref{ctr_insert_av} (\resp~\eqref{ctr_insert_retard})
discards every partition where $u$ is early (\resp tardy),
but which would have a lower penalty 
if $u$ were tardy (\resp early).
Note that 
the constraint $\Du \!\geqslant\! 0$ (\resp $\Du \!\leqslant\! 0$)
might be not satisfied by an optimal solution
where $u$ is tardy (\resp early).\\
Moreover,
only considering constraints 
\eqref{ctr_insert_av} and \eqref{ctr_insert_retard}
for a given $u$
is not sufficient to discard all insert-dominated partitions.
Indeed,
these constraints do not take into account the whole 
insert-neighborhood of the solutions as
each one is only compared to its neighbor
obtained by an insert operation on task $u$.
It is thus needed to consider
constraints~\eqref{ctr_insert_av} and~\eqref{ctr_insert_retard}
for every $u\iin J$
to translate the dominance of the insert local optimal solutions.
In Section~\ref{sec_dom_PL},
we will explain how these constraints
can be used in a linear formulation.

\subsubsection{Penalty variation induced by a swap operation}
\renewcommand {\Bu}{B(u)\!\setminus\!\{v\} \!\cap\!T}
\renewcommand {\Bv}{B(v)\!\cap\!T}

\noindent
This section follows the
same organization as the previous one.
The objective is to obtain constraints
translating the dominance of the swap locally optimal solutions.\\

Given $(u,v)\iin J^2$ such that $u\!\neq\!v$,
let $(E,T)$ be a partition such that $u\iin E$ and $v\iin T$,
and $\S$ be a $u$ and $v$-canonical 
representative of $(E,T)$.
We denote by $\S'$ the schedule obtained from $\S$ 
by swapping $u$ and $v$
so that $\S'$ is also both $u$ and $v$-canonical,
that is by scheduling $u$ 
after all the early tasks having the same $\a$-ratio $\a_u/p_u$
and $v$ before all the tardy tasks having the same $\b$-ratio $\b_v/p_v$.
$\S'$ is a representative of 
$\big(E\!\setminus\!\{u\}\!\cup\!\{v\},
T\!\setminus\!\{v\}\!\cup\!\{u\}\big)$.
Let $(e,t)$ (\resp $(e',t')$) denote
the earliness and tardiness vector
of tasks in $\S$ (\resp $\S'$).\\

If $ \frac{\a_v}{p_v} \!<\! \frac{\a_u}{p_u}$, 
all the early tasks of $\Ab(v)$ are 
scheduled before $u$ in $\S$,
since the \Vshaped property holds,
but the early tasks of $A(v)\!\setminus\!\{u\}$
can be scheduled before or after $u$.
This case is illustrated 
in Figure~\ref{fig_super_swap_av_Un}. 
Conversely,
if $\frac{\a_v}{p_v} \!\geqslant\! \frac{\a_u}{p_u}$,
all the early tasks of $A(v)$
are scheduled after $u$ in $\S$,
but those of \modifRun{$\Ab(v)\!\setminus\!\{u\}$}
can be scheduled before or after $u$.
This case is illustrated 
in Figure~\ref{fig_super_swap_av_Deux}.

\begin{figure}[h!]
\captionsetup{justification=centering}
\begin{minipage}{0.48\textwidth}
\centering
\OrdoCombAvUn{1}\caption{
Early side variation induced by swapping $u$ and $v$ 
when $\dfrac{\a_v}{p_v} \!<\! \dfrac{\a_u}{p_u}$
}
\label{fig_super_swap_av_Un}
\end{minipage}
\hspace{0.02\textwidth}
\begin{minipage}{0.48\textwidth}
\centering
\OrdoCombAvDeux{1}\caption{
Early side variation induced by swapping $u$ and $v$ 
when $\dfrac{\a_v}{p_v} \!\geqslant\! \dfrac{\a_u}{p_u}$
}
\label{fig_super_swap_av_Deux}
\end{minipage}
\end{figure}
\vspace{0.4cm}
\begin{figure}[h]
\captionsetup{justification=centering}
\begin{minipage}{0.48\textwidth}
\centering
\OrdoCombRetardUn{1}\caption{
Tardy side variation induced by swapping $u$ and $v$ 
when $\dfrac{\b_v}{p_v} \!\leqslant\! \dfrac{\b_u}{p_u}$ 
}
\label{fig_super_swap_retard_Un}
\end{minipage}
\hspace{0.02\textwidth}
\begin{minipage}{0.48\textwidth}
\centering
\OrdoCombRetardDeux{1}\caption{
Tardy side variation induced by swapping $u$ and $v$ 
when $\dfrac{\b_v}{p_v} \!>\! \dfrac{\b_u}{p_u}$
}
\label{fig_super_swap_retard_Deux}
\end{minipage}
\end{figure}

Similarly,
we can distinguish two cases
depending on 
the relative order of 
the $\b$-ratios of $u$ and $v$,
as illustrated in Figures~\ref{fig_super_swap_retard_Un}
and~\ref{fig_super_swap_retard_Deux}.

As shown in Figures~\ref{fig_super_swap_av_Un}
and~\ref{fig_super_swap_av_Deux},
the earliness of $u$ in $\S$ is
$e_u \!=\! p\big(\Au\big)$,
while the tardiness of $u$ in $\S'$ is 
$e'_u \!=\!p\big(\Bu\big) \+ p_u$
(\cf Figures~\ref{fig_super_swap_retard_Un}
and~\ref{fig_super_swap_retard_Deux}).
Note that $v$ is removed from $B(u)$
since $v$ is not tardy in $\S'$,
and therefore cannot contribute to the tardiness of $u$.
Similarly,
we have 
$t_v \!=\! p\big(\Bv\big) \+ p_v $ 
and $e'_v \!=\! p\big(\Avs\big)$ 
since $u$ is not early in $\S'$.\\[0.1cm]

Moreover,
the tasks of $\Au$
are identically scheduled in $\S$ and $\S'$
only if $\frac{\a_v}{p_v} \!<\! \frac{\a_u}{p_u}$.
In this case,
tasks of $\Abu$
are not consecutive in $\S'$
since $v$ separates them
into two blocks:
tasks of $\Abv$ which have been left-shifted
by $p_v\!-\!p_u$ time units,
and tasks of $\AAbvu$
which have been right-shifted 
by $p_u$ time units
(\cf Figure~\ref{fig_super_swap_av_Un}).
\\[0.1cm]
In the opposite case,
\ie if $\frac{\a_v}{p_v} \!\geqslant\! \frac{\a_u}{p_u}$,
tasks of $\Au$ are not consecutive in $\S'$.
Indeed, $v$ separates them into two blocks:
tasks of $\Av$
which are identically scheduled in $\S$ and $\S'$;
and tasks of $\AAbuv$
which are left-shifted by $p_v$ time units.
Moreover, in that case
tasks of $\Av$
are left-shifted by $p_v\-p_u$ time units
(\cf Figure~\ref{fig_super_swap_av_Deux}).
\\
Note that, 
in the two previous paragraphs,
as in Figures~\ref{fig_super_swap_av_Un}
to~\ref{fig_super_swap_retard_Deux},
we assumed that $p_v\!-\!p_u \!\geqslant\!0$.
In the contrary case,
\ie if $p_v\!-\!p_u \!<\!0$,
tasks are not left-shifted by $p_v\-p_u$ time units
but rather right-shifted by $p_u\-p_v$ time units.
\\

From these observations,
we can express the earliness penalty variation
of all the tasks in $E\!\setminus\!\{u\}$
by expressing, in each case,
the penalty variation induced by a block shifting as
described for the insert operation.
The same method
can be applied for the tasks of $T\!\setminus\!\{v\}$.
The reader can refer to
Figures~\ref{fig_super_swap_retard_Un}
and~\ref{fig_super_swap_retard_Deux}
for illustration.
Finally,
the penalty variation between $\S$ and $\S'$
is given by the following expression.
\begin{align*}
\Duv = 
&-\a_u \, p\big(\Au\big)
+\b_u \, \Big(p\big(\Bu\big) \+ p_u \Big)\\
&-\b_v \, \Big(p\big(\Bv\big) \+ p_v \Big)
+\a_v \,p\big(\Avs\big)\\[0.1cm]
&+\begin{cases}
(-p_u\+p_v)\, \a\big(\Abv\big) - p_u \, \a\big(\AAbvu\big)
&\text{if }\dfrac{\a_v}{p_v} \!<\! \dfrac{\a_u}{p_u}\\
(-p_u\+p_v)\, \a\big(\Abu\big) + p_v \, \a\big(\AAbuv\big)
&\text{otherwise}\\
\end{cases}\\[0.1cm]
&+\begin{cases}
(p_u\-p_v)\, \b\big(\Bbv\big) + p_u \, \b\big(\BBbvu\big)
&\text{if }\dfrac{\b_v}{p_v} \!\leqslant\! \dfrac{\b_u}{p_u}\\
(p_u\-p_v)\, \b\big(\Bbu\big) - p_v \, \b\big(\BBbuv\big)
&\text{otherwise}\\
\end{cases}
\end{align*}

\begin{pte}
Let $(E,T)$ be a partition.\\
For any $(u,v)\iin E\!\times\! T$,
$f\big(E\!\setminus\!\{u\}\!\cup\!\{v\},\, T\!\setminus\!\{v\}\!\cup\!\{u\}\big) 
= f(E,T) + \Duv$.
\label{pte_swap}
\end{pte}

\noindent
For a given pair of tasks $(u,v)\iin J^2$ 
such that $u \!\neq\! v$,
let us introduce the following constraint.
\begin{equation}
(u,v)\iin E\!\times\! T \Rightarrow\Duv \!\geqslant\! 0 \,
\tag{$S_{u,v}$}
\label{ctr_swap}
\end{equation}

\noindent
Thanks to Property~\ref{pte_swap},
\eqref{ctr_swap} discards 
every partition where $u$ is early, $v$ is tardy
and which would have a lower penalty 
in the contrary case.
As for the insert operation,
constraint $\Duv \!\geqslant\! 0$ 
might be not satisfied by optimal partitions
where $u$ is not early or $v$ is not tardy.
Moreover, it is needed to consider
constraints~\eqref{ctr_swap}
for every $(u,v)\iin J^2$ such that $u\!\neq\!v$
to translate the dominance of swap locally optimal solutions.
In Section~\ref{sec_dom_PL},
we will explain how these constraints
can be used in a linear formulation.

\section{Neighborhood based dominance in linear programming}\label{sec_dom_PL}
The dominance properties described in Section~\ref{sec_dom}
can be used in a linear programming framework. 
In this section,
we provide linear inequalities translating
constraints~\eqref{ctr_insert_av},~\eqref{ctr_insert_retard}
and~\eqref{ctr_swap} for all tasks $u$ and $v$.

\subsection{A compact MIP translating $\FET$}
We first recall the linear compact MIP formulation 
for the~\lepb
given in~\cite{nous19}.
In this formulation denoted by~$\FDXf$,
a partition is encoded by 
a vector $\delta$ of $n$ binary variables.
Given $\delta\iin\{0,1\}^J$,
the partition encoded by $\delta$ is
$\big(\{j\iin J \,|\, \delta_j \!=\!1\},
\{j\iin J \,|\, \delta_j \!=\!0\}\big)$.
In other words, 
for each $j\iin J$,
$\delta_j$ indicates whether $j$ is early or not.\\

\AF{variables de linearisation, inegalités}
Moreover,
$\FDXf$ also uses $X$,
a vector of binary variables indexed by the set 
$J^< = \{ (i,j)\iin J^2 \,|\, i<j\},$
in order to linearize 
products of type $\delta_i \,\delta_j$
as proposed by~\cite{Fortet}.

\begin{lm}[\cite{Fortet}]
Let $(\delta,X) \in \R^J \!\!\times\! \R^\J$.\\
\textbf{If} $\delta\iin\{0,1\}^J$ and 
$(\delta,X)$ satisfies 
the following inequalities:\\[-0.8cm]
\noindent \hspace*{-1cm}
\begin{minipage}[t]{0.52\textwidth}
\begin{eqnarray}
&&\forall (i,j) \!\in\! J^<,\enskip  X_{i,j} \geqslant \delta_i \!-\! \delta_j  
\label{x.1}\\
&&\forall (i,j) \!\in\! J^<,\enskip X_{i,j} \geqslant \delta_j \!-\! \delta_i 
\label{x.2}
\end{eqnarray}
\end{minipage}
\hspace*{-0.4cm}
\begin{minipage}[t]{0.58\textwidth}
\begin{eqnarray}
&&\forall (i,j) \!\in\! J^<,\enskip X_{i,j} \leqslant \delta_i \!+\! \delta_j 
\label{x.3}\\
&&\forall (i,j) \!\in\! J^<,\enskip X_{i,j} \leqslant 2 \!-\! (\delta_i \!+\! \delta_j)
\label{x.4}
\end{eqnarray}
\end{minipage} \\[0.2cm]
\textbf{then }
$\forall (i,j)\iin\J\!, X_{i,j}\=
\left\lbrace\begin{tabular}{@{}l@{ }l@{}}
1 & if $\delta_i \!\neq\!\delta_j$\\
0 & otherwise
\end{tabular}\right.$,
$\enskip(1\!-\!\delta_i) (1\!-\!\delta_j) \!=\! \dfrac{2\!-\!(\delta_i\!+\!\delta_j)\!-\!X_{i,j}}{2}$
and $ \delta_i\, \delta_j \!=\! 
\dfrac{\delta_i\!+\!\delta_j\!-\!X_{i,j}}{2}$.
\label{lm_x}
\end{lm}
Let us consider the polyhedron
$P \!=\! \left\{(\delta,X)\iin\R^J \!\!\times\! \R^{\J} 
\,\middle|\, 
(\ref{x.1}-\ref{x.4})\right\}$
and the set of its integer points
$\text{int}_\delta(P) \!=\! P \!\cap\! \{0,1\}^J \!\times\! \{0,1\}^{\J} $.
From Lemma~\ref{lm_x},
if a partition $(E,T)$ is encoded by $\delta$,
there exists a unique $X$ 
such that $(\delta,X)\iin \text{int}_\delta(P)$.
We will say that $(\delta,X)$ encodes $(E,T)$.
\\

\AF{intro de rho et sigma, fonction objectif}
To obtain an expression of the penalty of a partition 
from its encoding $(\delta,X)$,
two orders on $J$ are introduced.
Let $\rho$ and $\sigma$ be
two functions from $[1..n]$ to $J$,
such that
$$\left(\dfrac{\a_{\rho(k)}}{p_{\rho(k)}}\right)_{k\in[1..n]}
\text{ and }\enskip\enskip
\left(\dfrac{\b_{\sigma(k)}}{p_{\sigma(k)}}\right)_{k\in[1..n]}
\text{are non-increasing}
$$
A \dbloc is said \textit{\RSshaped} 
if the early (\resp tardy) tasks are processed in decreasing
order of $\rho^{-1}$ (resp. increasing order of $\sigma^{-1}$).
Each equivalence class of \Vdblocs admits
a unique \RSshaped representative.
This representative is used to provide
the following expression of the penalty
of the partition encoded by 
$(\delta,X)\iin\{0,1\}^J \!\!\times\! \{0,1\}^{\J}$.
\begin{align*}
g(\delta,X)\!= &\sum_{j\in J} 
\a_j\left(
\sum\limits_{k=1}^{\rho^{-1}(j) -1 } 
\hspace{-0.2cm} p_{\rho(k)} \, 
\dfrac{\delta_{j}\!+\!\delta_{\rho(k)}- X_{j,\rho(k)} }{2}
\right) \\
&+\, \b_j \left(
\sum\limits_{k=1}^{\sigma^{-1}(j) -1 } 
\hspace{-0.25cm}  p_{\sigma(k)} \,  
\dfrac{ 2 \!-\! (\delta_{j}\!+\!\delta_{\sigma(k)}) - X_{j,\sigma(k)} }{2}
+ p_j (1\!-\! \delta_j)\right)
\end{align*}

\AF{résumé de la formulation}
\noindent
Finally, 
the compact formulation for the \lepb 
provided in~\cite{nous19} is the following.
$$(\FDXf):\enskip{\min_{(\delta,X)\,\in\,\text{int}_\delta(P)} g(\delta,X)}$$
Formulation~$\FDXf$ 
is a direct linear translation of~$\FET$.
Indeed,
there is a one to one correspondence
between their solution sets,
\ie between $\text{int}_\delta(P)$
and  $\PartC$,
and $g(\delta,X)\=f(E,\,T)$
for any $(\delta,X)$ encoding $(E,T)$.

\subsection{Linear inequalities translating constraints~\eqref{ctr_insert_av}, \eqref{ctr_insert_retard} and~\eqref{ctr_swap}}

\subsubsection{Linear inequalities translating constraints~\eqref{ctr_insert_av} and \eqref{ctr_insert_retard} for any $u\iin J$}
\noindent
Let $u\iin J$.
If $\delta \iin\{0,1\}^J$
encodes a partition $(E,T)$,
the penalty variation $\Du$ 
can be expressed linearly from $\delta$ as follows.
$$
\Dud = -\a_u 
\hspace*{-0.1cm}\sum\limits_{i\in A(u)}\hspace*{-0.2cm} p_i\, \delta_i
+\b_u\,\Big(
\hspace*{-0.1cm}\sum\limits_{i\in B(u)}\hspace*{-0.2cm} p_i (1\!-\!\delta_i) 
\+ p_u \Big)
+ p_u\, \Big( 
\hspace*{-0.1cm}\sum\limits_{i\in \Bb(u)}\hspace*{-0.2cm}\b_i (1 \!-\!\delta_i)
-\hspace*{-0.1cm}\sum\limits_{i\in \Ab(u)}\hspace*{-0.2cm}\a_i\,\delta_i\Big)
$$

\noindent
Moreover,
if $\delta \iin\{0,1\}^J$ encodes $(E,T)$,
we can translate constraint~\eqref{ctr_insert_av}
as follows.\\[-0.8cm]
\begin{align*}
(E,T) \text{ satisfies~\eqref{ctr_insert_av}}
&\Leftrightarrow
\big(u\iin E \text{ and }\Du \!\geqslant\!0 \big) 
\text{ or } u\iin T\\
&\Leftrightarrow
\big(\delta_u\=1 \text{ and }\Du \!\geqslant\!0 \big) 
\text{ or } \delta_u\=0 
\end{align*}
To unify these two cases into one inequality,
we introduce the following constant
which is an upper bound of  
$-\Dud$ for any $\delta \iin\{0,1\}^J$.
$$\Mu \= 
\a_u \, p\big(A(u)\big)
-\b_u \, p_u
+ p_u\, \a\big(\Ab(u)\big)
$$
Since we have $\Dud \geqslant\! -\Mu$
for any $\delta \iin\{0,1\}^J$,
the following inequality is satisfied by 
every $\delta \iin\{0,1\}^J$ such that
$(1 \!-\!\delta_u) \!=\!1$.
\begin{equation}
\Dud \geqslant -\Mu (1 \!-\!\delta_u)
\label{ineg_insert_av}
\end{equation}
Conversely,
for every $\delta \iin\{0,1\}^J$ such that
$(1 \!-\!\delta_u) \=0$,
inequality~\eqref{ineg_insert_av} is satisfied if and only if 
$\Dud \!\geqslant\! 0$.\\
Considering these two cases,
$\delta \iin\{0,1\}^J$ satisfies~\eqref{ineg_insert_av}
if and only if the partition encoded by $\delta$
satisfies~\eqref{ctr_insert_av}.
Property~\ref{pte_insert_bis}.\textsl{(i)} follows.\\

Similarly,
to translate constraint~\eqref{ctr_insert_retard},
we introduce the following constant
which is an upper bound of  
$-\Dud$ for any $\delta \iin\{0,1\}^J$.
$$\Mpu \= 
\b_u \, p\big(B(u)\big)
+\b_u \, p_u
+ p_u\, \b\big(\Bb(u)\big)
$$
Since we have
$\forall\delta \iin\{0,1\}^J,\, -\Dud \geqslant\! -\Mpu$,
the following inequality is satisfied by 
every $\delta \iin\{0,1\}^J$ such that
$\delta_u \!=\!1$.
\begin{equation}
-\Dud \geqslant -\Mpu \,\delta_u
\label{ineg_insert_retard}
\end{equation}
Conversely, if $\delta_u \!=\!0$,
$\delta$ satisfies~\eqref{ineg_insert_retard}
if and only if $-\Dud \geqslant 0$.\\
Considering these two cases,
$\delta \iin\{0,1\}^J$
satisfies~\eqref{ineg_insert_retard}
if and only if 
if and only if the partition encoded by $\delta$
satisfies~\eqref{ctr_insert_av}.
Property~\ref{pte_insert_bis}.\textsl{(ii)} follows.


\begin{pte}
Let  $\delta \iin\{0,1\}^J$ and 
let $(E,T)$ be the partition encoded by $\delta$.
For any $u\iin J$,\\[0.1cm]
(i) $\delta$ satisfies inequality~\eqref{ineg_insert_av}
\textbf{if and only if} 
$(E,T)$ satisfies constraint~\eqref{ctr_insert_av}.
\\[0.1cm]
(ii) $\delta$ satisfies inequality~\eqref{ineg_insert_retard}
\textbf{if and only if} $(E,T)$ satisfies constraint~\eqref{ctr_insert_retard}.
\label{pte_insert_bis}
\end{pte}

\subsubsection{Linear inequalities translating constraint~\eqref{ctr_swap} for any $(u,v)\iin J^2$}
\noindent
Let $(u,v)\iin J^2$ such that $u\!\neq\!v$.
If $\delta \iin\{0,1\}^J$
encodes a partition $(E,T)$,
the penalty variation $\Duv$ 
can be expressed linearly from $\delta$ as follows.
\\[-0.4cm]
\begin{align*}
\hspace*{-0.9cm}\Duvd = 
&-\a_u \hspace*{-0.1cm}\sum\limits_{i\in A(u)}
\hspace*{-0.2cm} p_i\, \delta_i
+\b_u \hspace*{-0.4cm}\sum\limits_{i\in B(u)\setminus\{v\}}
\hspace*{-0.4cm} p_i\,(1\!-\!\delta_i) 
+\b_u p_u
-\b_v \hspace*{-0.2cm}\sum\limits_{i\in B(v)}
\hspace*{-0.2cm} p_i\,(1\!-\!\delta_i) 
-\b_v p_v
+\a_v \hspace*{-0.3cm}\sum\limits_{i\in A(v)\setminus\{u\}}
\hspace*{-0.4cm} p_i\, \delta_i
\\[0.3cm]
&+\begin{cases}
(-p_u\+p_v)\hspace*{-0.1cm}\sum\limits_{i\in \Ab(v)}
\hspace*{-0.2cm} \a_i\,\delta_i 
\hspace*{0.2cm} - \hspace*{0.2cm}
p_u \hspace*{-0.3cm}\sum\limits_{i\in A(v) \cap  \Ab(u) }
\hspace*{-0.3cm} \a_i\,\delta_i 
&\hspace*{1.5cm}\text{if }\dfrac{\a_v}{p_v} \!<\! \dfrac{\a_u}{p_u}\\[0.5cm]
(-p_u\+p_v)\hspace*{-0.1cm}\sum\limits_{i\in \Ab(u)}
\hspace*{-0.2cm} \a_i\,\delta_i 
\hspace*{0.2cm} + \hspace*{0.2cm}
p_v \hspace*{-0.3cm}\sum\limits_{i\in A(u) \cap  \Ab(v) }
\hspace*{-0.3cm} \a_i\,\delta_i 
&\hspace*{1.5cm}\text{otherwise}\\
\end{cases}\\[0.3cm]
&+\begin{cases}
(p_u\-p_v)\hspace*{-0.1cm}\sum\limits_{i\in \Bb(v)}
\hspace*{-0.2cm} \b_i\,(1\!-\!\delta_i) 
\hspace*{0.2cm} + \hspace*{0.2cm}
p_u \hspace*{-0.3cm}\sum\limits_{i\in B(v) \cap  \Bb(u) }
\hspace*{-0.3cm} \b_i\,(1\!-\!\delta_i)
&\text{if }\dfrac{\b_v}{p_v} \!\leqslant\!\dfrac{\b_u}{p_u}\\[0.5cm]
(p_u\-p_v)\hspace*{-0.1cm}\sum\limits_{i\in \Bb(u)}
\hspace*{-0.2cm} \b_i\,(1\!-\!\delta_i) 
\hspace*{0.2cm} - \hspace*{0.2cm}
p_v \hspace*{-0.3cm}\sum\limits_{i\in B(u) \cap  \Bb(v) }
\hspace*{-0.3cm} \b_i\,(1\!-\!\delta_i)
&\text{otherwise}\\
\end{cases}\\
\end{align*}

\noindent
Moreover,
if $\delta \iin\{0,1\}^J$ encodes $(E,T)$,
we can translate constraint~\eqref{ctr_swap}
as follows.\\[-0.8cm]
\begin{align*}
\hspace*{-0.8cm}(E,T)\text{ satisfies}~\eqref{ctr_swap}
&\Leftrightarrow
(u,v) \!\not\in\! E \!\times\! T 
\text{ or } \Duv \!\geqslant\! 0\\
&\Leftrightarrow
\big(u \!\not\in\! E 
\text{ or } v \!\not\in\! T \big)
\text{ or } \big( u\iin E 
\text{ and } v\iin T 
\text{ and } \Duv \!\geqslant\! 0 \big)\\
&\Leftrightarrow
\Big( (1\!-\!\delta_u) \!=\! 1 
\text{ or } \delta_v \!\=\! 1 \Big)
\text{ or } \Big(\big( (1\!-\!\delta_u) \!=\! 0
\text{ and } \delta_v \!\=\! 0\big)
\text{ and } \Duvd \!\geqslant\! 0 \Big)\\
&\Leftrightarrow
\big(\delta_v +(1 \-\delta_u) \big) \!\in\!  \{1,2\} 
\text{ or } \Big( \big(\delta_v +(1 \-\delta_u) \big)\=0
\text{ and } \Duvd \!\geqslant\! 0 \Big)
\end{align*}
To unify these cases into one inequality,
we introduce the following constant $\Mtuv$
which is an upper bound on
$-\Duvd$ for any $\delta \iin\{0,1\}^J$.
\\[-0.4cm]
\begin{align*}
\Mtuv = 
& \hspace*{0.2cm} \a_u \, p\big(A(u)\big)
-\b_u \, p_u
+\b_v \, p\big(B(v)\big)
+\b_v p_v\\[0.2cm]
&+\begin{cases}
[p_u\-p_v]^+\, \a\big(\Ab(v)\big)
+p_u \, \a\big(A(v) \!\cap\! \Ab(u)\big)
&\hspace*{0.2cm}\text{if }
\dfrac{\a_v}{p_v} \!<\! \dfrac{\a_u}{p_u}\\[0.4cm]
[p_u\-p_v]^+\, \a\big(\Ab(u)\big)
&\hspace*{0.2cm}\text{if }
\dfrac{\a_v}{p_v} \!\geqslant\! \dfrac{\a_u}{p_u}\\
\end{cases}\\[0.3cm]
&+\begin{cases}
[p_v\-p_u]^+\, \b\big(\Bb(v)\big)
&\text{if }\dfrac{\b_v}{p_v} \!\leqslant\!\dfrac{\b_u}{p_u}\\[0.4cm]
[p_v\-p_u]^+\, \b\big(\Bb(u)\big) 
+ p_v \, \b\big( B(u)\cap \Bb(v) \big) 
&\text{if }\dfrac{\b_v}{p_v} \!>\!\dfrac{\b_u}{p_u}\\
\end{cases}
\end{align*}

\noindent
Since we have $\Duvd \geqslant\! -\Mtuv$
for any $\delta \iin\{0,1\}^J$,
the inequality
$\Duvd \geqslant -\Mtuv \big(\delta_v +(1 \-\delta_u) \big)$
is satisfied for every $\delta \iin\{0,1\}^J$ 
such that $\big(\delta_v +(1 \-\delta_u) \big) \!=\!1$.
In particular,
this inequality is satisfied for 
every $\delta \iin\{0,1\}^J$ such that
$(\delta_u,\delta_v)\=(0,0)$
or $(1,1)$. 
For $\delta \iin\{0,1\}^J$ such that
$\big(\delta_v +(1 \-\delta_u) \big) \!=\!2$,
\ie such that $(\delta_u,\delta_v)\=(0,1)$,
this inequality
might not be satisfied
if $\Mtuv \!\leqslant\! 0$,
since $-\Mtuv \!\not\geqslant\! -2\Mtuv$
in this case.
To provide an inequality satisfied 
by every $\delta \iin\{0,1\}^J$ such that
$(\delta_u,\delta_v)\!\neq\!(1,0)$
we introduce the following constant.
$$\Muv=
\begin{cases}
\Mtuv &\text{ if } \Mtuv\!\geqslant\!0\\
\Mtuv/2 &\text{ otherwise}\\
\end{cases}
$$
We then have $-\Mtuv \!\geqslant\! -2\, \Muv$
and $-\Mtuv \!\geqslant -\Muv$.
Therefore,
the following inequality is satisfied by 
every $\delta \iin\{0,1\}^J$ such that
$\big(\delta_v +(1 \-\delta_u) \big)\iin\{1,2\}$,
\ie such that  $(\delta_u,\delta_v)\!\neq\!(1,0)$.
\begin{equation}
\Duvd \geqslant -\Muv \,\big(\delta_v +(1 \-\delta_u) \big)
\label{ineg_swap}
\end{equation}
Conversely,
for every $\big(\delta_v +(1 \-\delta_u) \big) \=0$,
$\delta \iin\{0,1\}^J$ such that
\ie such that $(\delta_u,\delta_v)\!=\!(1,0)$,
inequality~\eqref{ineg_swap} is satisfied if and only if 
$\Duvd \geqslant 0$.
Finally,
$\delta \iin\{0,1\}^J$ satisfies~\eqref{ineg_swap}
if and only if the partition encoded by $\delta$
satisfies~\eqref{ctr_swap}.
Property~\ref{pte_swap_bis} follows.

\begin{pte}
Let  $\delta \iin\{0,1\}^J$ and 
let $(E,T)$ be the partition encoded by $\delta$. 
For any $(u,v)\iin J^2$ such that $u\!\neq\!v$,
$\delta$ satisfies~\eqref{ineg_swap}
\textbf{if and only if} $(E,T)$ satisfies 
constraint~\eqref{ctr_swap}.\label{pte_swap_bis}
\end{pte}

In the sequel,
insert inequalities~\eqref{ineg_insert_av} 
and~\eqref{ineg_insert_retard}
and swap inequalities~\eqref{ineg_swap}
will be called \textit{dominance inequalities}.

\subsection{Link between dominance properties and operations}
In this section,
we show how to benefit from the fact that
each dominance inequality is based on an operation.
When a vector $\delta \iin\{0,1\}^J$
encoding a partition $(E,T)$
does not satisfy a dominance inequality,
applying the corresponding operation to $(E,T)$
provides a partition with a strictly lower penalty.
The following property,
resulting from properties~\ref{pte_insert},~\ref{pte_swap},
\ref{pte_insert_bis} and~\ref{pte_swap_bis},
formally states this result.

\begin{pte} 
Let  $\delta \iin\{0,1\}^J$ and 
let $(E,T)$ be the partition encoded by $\delta$. 
For any $u\iin J$,\\
(i) 
$\delta$ does not satisfy~\eqref{ineg_insert_av} for $u$
\textbf{if and only if} $u\iin E$ and 
$f\big(E\!\setminus\!\{u\},T\!\cup\!\{u\}\big)\!<\!f(E,T)$,
\\
(ii) 
$\delta$ does not satisfy~\eqref{ineg_insert_retard} for $u$
\textbf{if and only if} $u\iin T$ and 
$f\big(E\!\cup\!\{u\},T\!\setminus\!\{u\}\big)\!<\!f(E,T)$.
\\[0.1cm]
(iii) Moreover,
for any $(u,v)\iin J^2$ such that $u\!\neq\!v$, 
$\delta$ does not satisfy~\eqref{ineg_swap} for $(u,v)$
\textbf{if and only if} $(u,v)\iin E \!\times\! T$ and 
\hbox{$f\big( E \!\setminus\! \{u\} \!\cup\! \{v\},
T \!\setminus\! \{v\} \!\cup\!\{u\}\big)
\!<\! f(E,T)$}.\label{pte_amelioration}
\end{pte}

\begin{proof}
Let us fix $u\iin J$.
From Property~\ref{pte_insert_bis},
$\delta$ does not satisfy inequality~\eqref{ineg_insert_av}
if and only if 
$(E,T)$ does not satisfy constraints~\eqref{ctr_insert_av},
which is equivalent to
$u\iin E$ and $\Du \!<\! 0$.
Using Property~\ref{pte_insert},
it is equivalent to
$u\iin E$ and
$f\big( E \!\setminus\! \{u\} \!\cup\! \{v\},
T \!\setminus\! \{v\} \!\cup\!\{u\}\big)
\!-\! f(E,T) \!<\! 0$,
which proves \textsl{(i)}.
The proofs of \textsl{(ii)} and \textsl{(iii)}
follow the same scheme.
\end{proof}

Property~\ref{pte_amelioration}
will be used in Section~\ref{sec_expe}
to propose a local search procedure.
The following corollary,
which directly derives from the negation of statements 
\textsl{(i)}, \textsl{(ii)} and \textsl{(iii)},
ensures that the solution provided by this local search procedure
is an insert and swap local optimum.

\begin{cor}
Let  $\delta \iin\{0,1\}^J$ and 
let $(E,T)$ be the partition encoded by $\delta$.\\[0.1cm]
(i) $(E,T)$ is an insert local optimum
\textbf{if and only if}
$\delta$ satisfies inequalities~\eqref{ineg_insert_av} 
and~\eqref{ineg_insert_retard} for all $u\iin J$.\\
(ii) $(E,T)$ is a swap local optimum
\textbf{if and only if}
$\delta$ satisfies inequality~\eqref{ineg_swap} 
for all $(u,v)\iin J^2$ such that $u\!\neq\!v$.
\label{corrolaire}
\end{cor}


The following section presents experimental results
to assess the practical relevance of the dominance inequalities. 
\section{Numerical results}
\label{sec_expe}
All experiments are carried out using a single thread with 
Intel(R) Xeon(R) X5677, @ 3.47GHz,
and 144Gb RAM.
Linear programs (LP) and MIP 
are solved with Cplex 12.6.3.0.\\

The numerical experiments are performed
on the instance benchmark
proposed by~\cite{benchmark},
available online on OR-Library~\citep{benchmark_url}.
For each number of tasks 
$n \!\in\! \{10,20,50,100,200\}$,
ten triples $(p,\alpha,\beta)$ of $\big({\NNe}^n\big)^3$ are given.
For each one,
we assume that $d \!=\! p(J)$, 
so that the due date is unrestrictive.
For the sake of comparison,
we additionally construct instances with $n\iin\{60,80\}$
(\resp $n\iin\{120,150,180\}$)
by only considering the first $n$ tasks 
of the previous $100$-task (\resp $200$-task) instances.
Unless otherwise specified,
the gap, time and number of nodes presented 
in the following tables
are average values over the ten instances
for a given $n$
and the time limit is set to 3600 seconds.\\

To measure the improvement induced by
the insert or swap inequalities,
we compare the four following formulations.
\newpage

\noindent
\renewcommand{\arraystretch}{1.2}
\begin{tabular}{r@{ : }p{14.5cm}}
$\FDXf$& the formulation defined in Section~\ref{sec_dom_PL},
only with inequalities (\ref{x.1}-\ref{x.4})\\
$\FIf$& the formulation obtained from $\FDXf$ 
by adding~\eqref{ineg_insert_av}
and~\eqref{ineg_insert_retard} for all $u\iin J$\\
$\FSf$&the formulation obtained from $\FDXf$ 
by adding~\eqref{ineg_swap}
for all $(u,v)\iin J^2 \text{ s.t. } u\!\neq\!v$\\
$\FISf$&the formulation obtained from $\FSf$ 
by adding~\eqref{ineg_insert_av}
and~\eqref{ineg_insert_retard} for all $u\iin J$
\end{tabular}\\
\renewcommand{\arraystretch}{1}

For a given formulation $F$,
we distinguish two settings:
a setting with all available features,
that is using Cplex default, denoted by $F_d$,
and a setting with less Cplex features, denoted by $F_l$.
Two types of features 
are disabled in this setting:
the cut generation,
which produces reinforcement inequalities and
adds them to the formulation,
and the primal heuristic procedures.
The cut generation is disabled 
in order to measure the impact of the dominance inequalities
on the linear relaxation value of the formulation $\FDXf$,
rather than their impact
on the linear relaxation value 
of a strengthened formulation.
The primal heuristic procedures have been disabled
to focus on the lower bound since we have other methods
to quickly obtain good feasible solutions
(\cf Section~\ref{subsec_UB}).

This results in eight formulation settings:
$\FDX,\FI,\FS,\FIS,\FF,\FFI,\FFS$ and $\FFIS$.
For each one,
inequalities~{(\ref{x.1}-\ref{x.4})},
as well as 
inequalities~(\ref{ineg_insert_av}-\ref{ineg_swap})
when included,
are added initially.\\

Let us recall that only the $\delta$ variables
need to be integer in $\FDXf$.
Indeed,
from Lemma~\ref{lm_x},
if $\delta \iin\{0,1\}^J$,
inequalities (\ref{x.1}-\ref{x.4})
ensures that $X\iin\{0,1\}^\J$.
It is also the case for $\FIf,\FSf$ and $\FISf$.
Therefore,
unless otherwise specified,
$\delta$ variables are set as binary variables, while
$X$ variables are set as continuous variables.
Consequently, the branching decisions 
only involve $\delta$.

\subsection{Solving  MIP formulations to optimality}
\AF{présentation table 1}
Table~\ref{tbl_exact} provides the results
obtained by solving MIP to optimality,
using the eight formulation settings.
Each line corresponds to the ten instances
of a given size $n$.
More precisely,
Table~\ref{tbl_exact} entries
are the following.

\noindent
\begin{tabular}{r@{ : }p{11.5cm}}
\#opt & the number of instances solved to optimality 
within the time limit\\
\time & the average running time in seconds 
over the instances solved to optimality\\
\nd & the average number of nodes, except the root node,
in the search tree, 
over the instances solved to optimality\\
\end{tabular}\\
For a given formulation setting,
we choose to stop the run at a line of the table
if less than 5 over ten instances are solved to optimality.
For the subsequent lines,
we report a "-" in the table.\\

\AF{Description/Conclusions table 1}
Using formulation setting $\FDX$,
the ten $n$-task instances are solved to optimality
within the time limit for $n$ up to $50$.
In contrast,
using $\FI$,
it is the case for $n$ up to $60$,
using $\FS$ for $n$ up to $120$ 
and using $\FIS$ for $n$ up to $150$. 
Within approximately 5 minutes,
$\FDX$ solves $50$-task instances,
$\FI$ $60$-task instances,
$\FS$ $100$-task instances
and $\FIS$ $120$-task instances.
This computation time decrease is 
due to a drastic reduction in the number of nodes.
For example, for $n\!=\!50$ 
the number of nodes goes from 
more than 53 000 for $\FDX$
to only $31$ for $\FIS$.
With this latter formulation setting,
the number of nodes is low, it is at most 200,
even for large size instances.
However,
the time limit is reached 
for some $180$- and $200$-task instances,
since the size of the linear program solved at each node 
is large.\\

In light of the four first columns 
of Table~\ref{tbl_exact},
we can conclude that, 
with less Cplex features,
adding insert and swap inequalities 
significantly reduces  
the number of nodes and hence the computation time.
More precisely,
adding only swap inequalities 
is better than adding only insert inequalities,
but adding both of them provides the best performance.

The four last columns show the same improvement
in terms of computation time and number of nodes
when \modifRun{Cplex default features} are used.\\

Let us now focus on the 4th and 5th columns 
to compare the impact of the dominance inequalities 
and the impact of \modifRun{Cplex default features}.
For small instances, \ie $n\iin\{10,20\}$,
\modifRun{Cplex default features} allow to solve the problem
at the root node (\cf $\FF$ columns).
However,
from $n\!=\!50$, 
the number of nodes grows fast,
so that no 60-task instance can be solved
within 3600 seconds.
Conversely, 
we already noticed that
adding swap and insert inequalities
limits the number of nodes (\cf $\FIS$ columns),
so that the ten $150$-task instances are solved
within 3600 seconds.
Finally,
adding the swap and insert inequalities
provides better results than
adding \modifRun{Cplex default features}.\\

\modifRun{Up to size 60,
$\FFIS$ solves all instances at the root node,
and is faster than $\FIS$.
}
For larger instances,
except 200-task instances,
$\FIS$ and $\FFIS$ solves the problem 
in similar computation times,
even if $\FIS$ \modifRun{explores} a smaller number of nodes:
for example,
it is two times smaller for $n\=150$.
For $n\=200$,
$\FFIS$ solves 4 over 10 instances, 
while $\FIS$ only solves 1 over 10 instances.
To conclude, 
$\FIS$ and $\FFIS$ offer comparable performances,
\modifRun{
so that, for both settings, 
the formulations providing the best results 
are the ones with insertion and swap inequalities.
}

\subsection{Lower bound obtained at the root node}
\label{sec_expe_LB}

\AF{présentation table 2}
To further investigate the impact of dominance inequalities,
we focus in this section on 
the root node of the search tree
for different formulation settings.
More precisely,
we compare the different lower bounds 
obtained at the root node.\\

In the Cplex framework,
setting the node limit to 0 allows 
to only solve the root node of a MIP:
the branch-and-bound algorithm is stopped
before the first branching.
If \modifRun{Cplex default features} are activated,
the preprocessing is applied
and the cuts are added before the algorithm stops.
For a given formulation setting $F$,
the corresponding run with the node limit set to 0
is denoted by $\FRN$.
This results in eight runs:
$\FDXRN,\FIRN,\FSRN,\FISRN,\FFRN,\FFIRN,\FFSRN,\FFISRN$.
\\

\noindent
Note that, in the Cplex framework,
solving $\FRN$
is different from 
solving the linear relaxation of $F$,
denoted by $\FLP$.
Indeed, 
$\FLP$ is obtained by setting $\delta$ variables 
as continuous variables,
which desactivates most of the Cplex features.
In particular,
the reinforcement cuts cannot be added
since they are not valid for the relaxed formulation.
Similarly,
the inference procedure on the binary variables
cannot be applied.
We run the four linear relaxations 
$\FDXLP, \FILP, \FSLP$ and $\FISLP$.
Surprisingly,
the obtained values are the same 
for these four relaxations.
In other words,
adding insert and swap inequalities
does not improve the linear relaxation value.
Therefore,
we only present in Table~\ref{tbl_lowerB} 
the results for $\FDXLP$.\\

To measure the quality of the nine different lower bounds obtained,
we compute, 
when it is possible,
the optimality gap, 
\ie $(OPT\-LB)/OPT$ 
where $OPT$ denotes the optimal value
and $LB$ the lower bound.
When the optimal value is not known,
we compute a gap using the best upper bound that we get $UB$,
\ie $(UB\-LB)/UB$.
\modifAEF{Such} gaps are indicated with a "*" 
in Table~\ref{tbl_lowerB}.
For each of the nine runs,
the entries of Table~\ref{tbl_lowerB} 
are the following.\\[0.2cm]
\begin{tabular}{r@{ : }p{16cm}}
\LB & the average optimality gap of the lower bound 
obtained at the root node\\
\time& the average running time in seconds 
over the ten instances\\
\end{tabular}\\

\AF{description et conclusions table 2}
The obtained lower bound is exactly the same
using $\FDXLP$, $\FDXRN$ or $\FSRN$.
We deduce that with less Cplex features
and without insert inequalities,
setting the $\delta$ variables 
as binary or continuous variables,
provides the same lower bound.
Moreover,
this lower bound is quite weak,
since the average optimality gap
is larger than 40\% even for the 10-task instances.
The computation times using $\FDXLP$ and $\FDXRN$ 
are similar: 
2 seconds for the $100$-task instances
and about 12 minutes for the $500$-task instances.
The computation time required for $\FSRN$ is larger:
almost 20 seconds for the $100$-task instances
and 47 minutes for the $500$-task instances.\\

The lower bound obtained when 
only considering the insert inequalities
is slightly better  
when the $\delta$ variables are
set as binary variables
for $n \iin \{10,20\}$.
Indeed,
the average optimality gap is 
33\% instead of 41\% when $n\=10$,
and 66\% instead of 68\% when $n\=20$
(\cf $\FDXLP$ and $\FIRN $ columns).
The computation time using $\FIRN$ 
is comparable to the computation time 
using $\FDXLP$ and $\FDXRN$.\\

The lower bound obtained when
considering both insert and swap inequalities,
is significantly better 
when the $\delta$ variables are
set as binary variables.
Indeed,
the average optimality gap is 
smaller than 39\% for any value of $n$,
and  it is equal to 0 for $n\=10$
(\cf $\FISRN$ column).
The computation time for $\FISRN$ 
is between those for $\FIRN$ and $\FSRN$:
14 seconds for the $100$-task instances
and about 30 minutes for the $500$-task instances.
\\

%

\noindent
The lower bound provided using $\FFRN$,
is better than the one obtained using $\FDXRN$,
that is with less Cplex features.
Indeed,
the average optimality gap is 
7\% instead of 41\% when $n\=10$,
and 46\% instead of 94\% when $n\=120$.
However,
the lower bound is weaker
that the one obtained for $\FIS$,
whose optimality gap is
0 for $n\=10$ and 
38\% for $n\=120$.
Moreover,
the computation times using $\FFRN$
increases fast with the increase of $n$
so that 
the root node cannot be solved within one hour
for sizes larger than \modifAEF{120}.\\

Combining Cplex features with insert inequalities
gives almost the same results
(\cf $\FFIRN$ column).
Conversely,
combining Cplex features with swap inequalities
gives better results
(\cf $\FFSRN$ column).
In particular, the average computation time 
is reduced so that instances up to size 200
can be solved.
Moreover,
the optimality gap is less than 22\%
for all solved instances.
Finally,
using $\FFISRN$
gives even better results,
the average optimality gap does not exceed 15\%,
even for 200-task instances,
which are solved in 418 seconds,
instead of 1200 using $\FFSRN$.\\


%

In a nutshell,
combining insert and swap inequalities 
is the best
to obtain a lower bound at the root node.
Not using Cplex features allows
its fast computation
(\cf $\FISRN$ column).
Conversely,
using them allows
to obtain a better lower bound 
at the expense of the computation time
(\cf $\FFISRN$ column).

\subsection{Using swap and insert inequalities
to obtain an upper bound}
\label{subsec_UB}

\AF{petite intro}
In this section,
we propose two upper bounds on the optimal value.
The first one is derived from the fractional solution obtained at the root node by a simple rounding procedure.
The second one is obtained by applying in addition
a local search procedure.\\[-0.2cm]

\AF{introduire le rounding}
\noindent
We derive an integer solution $(\delta,X)$ 
by rounding a fractional solution 
$(\widetilde{\delta},\tilde{X)}$,
as follows.\\

$$\forall j \iin J,\,
\delta_j \= \begin{cases}
0
\text{ if  } \widetilde{\delta_j} \!<\! \frac{1}{2} 
\text{ or } \big(\widetilde{\delta_j} \!=\! \frac{1}{2}  
\text{ and } \a_j \!<\! \b_j\big)\\
1 \text{ otherwise}
\end{cases}
\hspace*{-0.3cm}\text{and } 
\forall (i,j)\iin \J\!,\,
X_{i,j}\=
\left\lbrace\begin{tabular}{@{}l@{ }l@{}}
1 & if $\delta_i \!\neq\! \delta_j$\\
0 & otherwise
\end{tabular}\right.
$$
By construction, 
$(\delta,X)$ satisfies inequalities~(\ref{x.1}-\ref{x.4})
(\cf Lemma~\ref{lm_x}).
It is thus a solution of $\FDXf$,
and $g(\delta,X)$ is an upper bound of the optimal value.
However,
it is not necessarily a solution 
for  $\FIf$, $\FSf$ and $\FISf$ formulations,
since \modifRun{$(\delta,X)$} does not necessarily satisfy
the insert and swap inequalities.\\

\AF{Introduire le rounding +}
In order to transform such a solution into a solution
satisfying the dominance inequalities,
we can iteratively apply the operation associated 
to each violated dominance inequality,
until all of them are satisfied.
Algorithm 1 presents a way to implement this procedure
that we call \LSproc.
From Property~\ref{pte_amelioration},
if an insert (\resp a swap) inequality
is not satisfied,
applying the appropriate insert (\resp swap) operation
provides a strictly better solution.
\modifRtrois{
Therefore, each solution is considered 
at most once in this procedure.
Since the number of solutions is finte, 
the \LSproc procedure finishes.
}

The returned solution is an insert and swap local optimum,
since it satisfies all dominance inequalities
(\cf Corrolary~\ref{corrolaire}).\\[-0.4cm]

\begin{center}
\begin{minipage}{0.95 \textwidth}
\newcommand{\tab}{\hspace*{0.5cm}}
\newcommand{\bool}{is\_locally\_opt\xspace}
\noindent
\LSproc\\[-0.2cm]
\hrule \vspace*{0.2cm}
\textbf{input}: $\delta \iin \{0,1\}^J$\\
\textbf{output}: $\delta'$ encoding 
an insert and swap local optimum\\[-0.2cm]
\hrule\vspace*{0.2cm}
\texttt{
\noindent
\tab \hspace*{-0.2cm}$\delta' \leftarrow \delta$;
\bool $\leftarrow$ false\\
\tab while (not \bool)\\
\tab\tab \bool $\leftarrow$ true\\
\tab \tab for $u\iin J$\\
\tab \tab \tab 
\modifRtrois{if $\delta'_u \!=\! 1$ and $\Delta_u(\delta') \!<\!0$} 
\textrm{\small \, //$\delta'$ does not satisfy \eqref{ineg_insert_av}}
\\
\tab\tab\tab\tab 
$\delta'_u \leftarrow\!$ 0;
\bool $\leftarrow$ false\\[0.1cm]
\tab \tab \tab 
\modifRtrois{if $\delta'_u \!=\! 0$ and $\Delta_u(\delta') \!>\!0$}
\textrm{\small \, //$\delta'$ does not satisfy \eqref{ineg_insert_retard}}
\\
\tab\tab\tab\tab 
$\delta'_u \leftarrow\!$ 1;
\bool $\leftarrow$ false\\
\tab \tab \tab for $v\iin J \!\setminus\! \{u\}$\\
\tab \tab \tab \tab
\modifRtrois{if $\delta'_u \!=\! 1$, $\delta'_v \!=\! 0$ and $\Delta_{u,v}(\delta') \!<\!0$}
\textrm{\small \, //$\delta'$ does not satisfy \eqref{ineg_swap}}
 \\
\tab \tab \tab \tab \tab 
$\delta'_u  \leftarrow\!$ 0;
$\delta'_v \leftarrow\!$ 1;
\bool $\leftarrow$ false\\[0.1cm]
\tab return $\delta'$\\[-0.2cm]
\hrule
}
\end{minipage}\\[0.4cm]
Algorithm 1: 
the improvement procedure by insert and swap operations\\
\end{center}

\noindent
Note that this algorithm 
can be seen as a local search procedure
for the neighborhood associated to
the insert and swap operations.
Moreover,
this procedure can be applied to any integer solution.
Particularly,
by sake of comparison
we apply it to the solutions obtained
by the heuristic "Heur II" provided
\modifAEF{by~\cite{benchmark}}.\\

\AF{présenter table 3}
\noindent
We finally compare the upper bounds
given by the six following heuristic solutions.\\[0.2cm]
\renewcommand{\arraystretch}{1.2}
\begin{tabular}{r@{ : }p{15cm}}
$\HBF$
& the solution obtained by 
the Biskup and Feldmann heuristic\\
$\HBFp$
& the solution obtained by 
applying \LSproc to $\HBF$\\
$\HR$
& the solution obtained by rounding
the fractional solution of $\FDXLP$\\
$\HRp$
& the solution obtained by 
applying \LSproc to $\HR$\\
$\HRR$
& the solution obtained by rounding
the fractional solution of $\FFISRN$\\
$\HRRp$
& the solution obtained by 
applying \LSproc to $\HRR$
\end{tabular}\\
\renewcommand{\arraystretch}{1}

In the sequel,
we will use the same notation for
both a heuristic solution and its value,
which provides an upper bound on the optimal value.
To measure the quality of these upper bounds, 
Table~\ref{tbl_upperB} presents 
their optimality gap denoted by \UB,
\ie $(UB-OPT)/OPT$ 
where $OPT$ denotes the optimal value
and $UB$ the upper bound.
The Biskup and Feldmann heuristic
provides a solution in less than 1 second.
Applying rounding and \LSproc
to a fractional solution provides 
a solution in less than 1 second 
for instances up to size 200.
Therefore,
the time needed to obtain $\HR$ and $\HRp$ 
(\resp $\HRR$ and $\HRRp$)
is essentially the computation time 
required to solve $\FDXLP$ (\resp $\FFISRN$) 
given in Table~\ref{tbl_lowerB}.\\

\AF{description table 3}
As shown in Table~\ref{tbl_upperB},
$\HBF$ is a good upper bound.
Indeed, its optimality gap is smaller than 0.35\%
for instance sizes larger than 50.
However,
this bound is improved by \LSproc:
the optimality gap of $\HBFp$
is smaller than 0.02\% for all the instances.
With an optimality gap larger than 170\%,
$\HR$ is a very weak upper bound,
while $\HRp$,
with an optimality gap smaller than 0.01\%,
is very good,
and even slightly better than $\HBFp$.
With an optimality gap smaller than 17\%,
$\HRR$ is a better upper bound than $\HR$,
and $\HRRp$ is exactly the same as $\HRp$.\\

\AF{conclusions table 3}
Finally, $\HBFp,\HRp$ and $\HRRp$
are very good upper bounds.
However it is worth noticing that 
even if the computation time to obtain $\HBFp$ 
is about 1 second,
the bound is obtained without any guarantee,
since no lower bound is provided.
Conversely,
the computation time to obtain $\HRRp$ 
is larger: 
25 seconds for $n \!=\! 100$
and about \modifAEF{7} minutes for $n \!=\! 200$,
but a lower bound is provided.
$\HRRp$ is then guaranteed 
to be at 14\% of the optimal value  for $n \!=\! 100$,
and at 15\% for $n \!=\! 200$
(\cf \LB of $\FFISRN$ in Table~\ref{tbl_lowerB}).
$\HRp$ is a compromise 
between $\HBFp$ and $\HRRp$.
Indeed,
for instances up to size 200,
$\HRp$ is provided in less than 20 seconds 
together with a lower bound,
but the guarantee obtained from
this lower bound is quite weak
(97\% for $n\=200$,
\cf $\FDXRN$ in Table~\ref{tbl_lowerB}).

\subsection{Insert and swap operations use cases}
\AF{présentation table 4}
Insert and swap operations can be used
in different ways.
Table~\ref{tbl_ccl} presents the best way to use them
depending on the expected solution quality.
\begin{itemize}
\item[-]
To obtain an upper bound:
apply rounding and \LSproc
to the fractional solution given by $\FDXLP$.
(\cf \FA column in Table~\ref{tbl_ccl}).
\item[-]
To obtain an upper bound with 
a better guarantee than the one obtained with \FA:
apply rounding and \LSproc
to the fractional solution given by $\FFISRN$.
(\cf \FB column in Table~\ref{tbl_ccl}).
\item[-]
To obtain a 5\%-approached solution: use $\FFIS$,
setting the gap limit to 5\%.
(\cf \FC column in Table~\ref{tbl_ccl}).
\item[-]
To obtain an exact solution: use $\FFIS$.
(\cf \FD column in Table~\ref{tbl_ccl}).
\end{itemize}

\noindent
Table~\ref{tbl_ccl} sums up the performance 
of the four above mentioned use  cases.
To measure the performances on the 200-task instances,
no time limit is fixed.
The entries of Table~\ref{tbl_ccl}
are the following.
\\[0.2cm]
\begin{tabular}{r@{ : }p{16cm}}
\LB & 
the average optimality gap of the provided lower bound\\
\UB & 
the average optimality gap of the provided upper bound\\
\time & the average running time in seconds\\
\nd & the average number of nodes except the root node \\
\end{tabular}\\

\AF{ce qui se répète/est nouveau dans cette table}
New experiments are conducted 
for the results reported in \FC and \FD columns 
when $n\!=\!200$.
These results are gathered with
the previously obtained results 
in Table~\ref{tbl_ccl}
to offer an overview.
\\

\AF{conclusions table 4}
\modifRun{
Table~\ref{tbl_ccl} shows that the number of nodes is lowered by 37.0\%
while the computation time  is only lowered  by 10.8\% on average for $n\=200$.
}
In addition, 
for the six 200-task instances
where $\FFIS$ reaches the time limit,
only less than 100 nodes are explored.
The limit for solving $\FIS$
is thus the size or the difficulty 
of the LPs solved at each node,
rather than the number of nodes.\\

\AF{Séparation et inférence}
Trying to address this issue,
we implemented a separation algorithm 
for the insert and swap inequalities
using a callback function.
The time needed to solve 50-task instances
using this separation algorithm
and Cplex features was 1513 seconds with 925 nodes in average.
We observe that 98\% of the computation time
is used by the UserCut Callback 
to add 71 inequalities in average.
This is not surprising since 
the separation algorithm consists in simply
evaluating the terms of 
inequality~\eqref{ineg_insert_av} and \eqref{ineg_insert_retard}
for the $n$ possible tasks $u$,
and the terms of inequality~\eqref{ineg_swap}
for the $n^2$ possible couples $(u,v)$,
which results in an $O(n^3)$ procedure.\\

Providing a faster separation algorithm
could reduce the computation time,
but the branching scheme, 
and then the number of nodes,
would be the same.
Since this number of nodes is quite large
compared to the performance of $\FFIS$
(which solves all 50-task instances at the root node),
we conclude that adding dominance inequalities
through a separation procedure reduces their impact.\\
Indeed,
when initially added,
the dominance inequalities allow to
the Cplex presolve phase to
fix some variables to 0 or 1.
The number of LPs variables is then reduced
and the value obtained at each node is improved.
When the $\delta$ variables are set as continuous variables,
this presolve is not executed.
It is then consistent with the observation
that adding dominance inequalities
in this latter case has no impact
(\cf Section~\ref{sec_expe_LB}).

\section{Conclusion}
In this work,
we propose a new way to use 
neighborhood-based dominance properties,
which results in a new kind of reinforcement inequalities.
In contrast with 
the commonly used reinforcement inequalities,
which cut fractional points,
these inequalities cut non locally optimal solutions.
In particular,
for the compact formulation $\FDXf$,
we provide linear inequalities
cutting all the solutions which are not
insert and swap locally optimal.

From a practical point of view,
we show that adding 
insert and swap inequalities
greatly improves performances of $\FDXf$.
Indeed, instead of $50$-task instances,
we can now solve up to $150$-task instances
to optimality within one hour.

Insert and swap inequalities can also be used 
to provide a heuristic solution
which is slightly better than the one proposed
by~\cite{benchmark}.
For instances up to size 200,
this heuristic solution is obtained 
in less than 20 seconds.
A lower bound \modifRun{providing} a 15\% gap
can also be obtained
in less than 420 seconds.\\

We observe that insert and swap inequalities 
do not improve the linear relaxation value 
of the compact formulation $\FDXf$.
However, 
used in conjunction with Cplex features,
they allow to improve the lower bound
obtained at the root node.
Two issues follow.
Firstly,
for a version of $\FDXf$
reinforced by cuts or by branching decisions,
do dominance properties improve 
the linear relaxation value?
Secondly,
which procedure implemented in the Cplex features
take advantage of the insert and swap inequalities?
Addressing these issues requires an appropriate experimental framework.\\

Moreover,
this work could be extended to other problems
where the solutions can be encoded by partitions
(any kind of partition,
not necessarily ordered bi-partitions).
For instance,
inequalities similar to insert and swap inequalities
could be used in a generalization of \lepb 
to a parallel machine framework.
Indeed,
if the tasks share a common due date,
the dominant schedules can be 
encoded by ordered $2m$-partitions,
where $m$ is the number of machines.  
This is true even if the common due date,
the processing times
and the unit earliness and tardiness penalties 
depend on the machine.
Beyond the scheduling field,
such inequalities
could also be used in the maximum cut problem~\citep{Karp_72}
or in a maximum $k$-cut problem~\citep{approx_max_k_cut_97}.

For other combinatorial problems
where solutions do not have a partition structure;
some neighborhood-based dominance inequalities 
could also be designed using 
appropriate operations.




\bibliography{biblio_restreinte_article2}

\begin{landscape}
\begin{table}
\small
\centering
\renewcommand{\arraystretch}{1.3}
\vspace*{-2cm}
\hspace*{-2cm}
\tblExact
\caption{Effect of adding insert and swap inequalities 
on exact solving }
\label{tbl_exact}

\vspace*{0.5cm}

\hspace*{-0.8cm}
\tblLB
\caption{Effect of adding insert and swap inequalities on lower bounds }
\label{tbl_lowerB}

\footnotetext{
NB: gap followed by a star are computed
using an upper bound instead of the optimal value,
since this optimal value is not known
}

\end{table}

\begin{table}
\centering
\renewcommand{\arraystretch}{1.3}

\tblUBvUn

\caption{Comparison of different heuristics
providing an upper bound}
\label{tbl_upperB}

\vspace*{2cm}
\tblCcl
\caption{Different ways of using
insert and swap inequalities}
\label{tbl_ccl}

\end{table}
\end{landscape}

\end{document}